\documentclass[sigconf,anonymous=False,review=False]{acmart}

\usepackage{booktabs} 
\usepackage{lipsum} 
\usepackage{balance}
\usepackage{paralist}
\usepackage{filecontents}
\usepackage{etoolbox}
\usepackage{mathrsfs}
\usepackage{bm}
\usepackage{setspace}
\usepackage{graphicx}
\usepackage{caption}
\usepackage{subcaption}
\usepackage{multirow}
\setcopyright{rightsretained}

\usepackage[utf8]{inputenc}
\usepackage[english]{babel}

\begin{document}
\title{Model-agnostic vs. Model-intrinsic Interpretability for Explainable Product Search}

\copyrightyear{2021} 
\acmYear{2021} 
\setcopyright{acmcopyright}\acmConference[CIKM '21]{Proceedings of the 30th ACM International Conference on Information and Knowledge Management}{November 1--5, 2021}{Virtual Event, QLD, Australia}
\acmBooktitle{Proceedings of the 30th ACM International Conference on Information and Knowledge Management (CIKM '21), November 1--5, 2021, Virtual Event, QLD, Australia}
\acmPrice{15.00}
\acmDOI{10.1145/3459637.3482276}
\acmISBN{978-1-4503-8446-9/21/11}

\fancyhead{}

\author{Qingyao Ai}
\authornote{Equal Contribution}
\affiliation{%
	\institution{University of Utah}
	\city{Salt Lake City} 
	\state{UT} 
	\country{USA}
}
\email{aiqy@cs.utah.edu}

\author{Lakshmi Narayanan Ramasamy}
\authornotemark[1]
\affiliation{%
	\institution{University of Utah}
	\city{Salt Lake City} 
	\state{UT} 
	\country{USA}
}
\email{lakshminarayanan.ramasamy@utah.edu}

\begin{abstract}

Product retrieval systems have served as the main entry for customers to discover and purchase products online.
With increasing concerns on the transparency and accountability of AI systems, studies on explainable information retrieval has received more and more attention in the research community.
Interestingly, in the domain of e-commerce, despite the extensive studies on explainable product recommendation, the studies of explainable product search is still in an early stage.
In this paper, we study how to construct effective explainable product search by comparing model-agnostic explanation paradigms with model-intrinsic paradigms and analyzing the important factors that determine the performance of product search explanations.
We propose an explainable product search model with model-intrinsic interpretability and conduct crowdsourcing to compare it with the state-of-the-art explainable product search model with model-agnostic interpretability.
We observe that both paradigms have their own advantages and the effectiveness of search explanations on different properties are affected by different factors.
For example, explanation fidelity is more important for user's overall satisfaction on the system while explanation novelty may be more useful in attracting user purchases.
These findings could have important implications for the future studies and design of explainable product search engines.
 
\end{abstract}

%
%
\iftrue

\fi


\maketitle

\section{Introduction}

As online marketplaces have gradually dominated the retail market, product retrieval systems such as product search engines and recommendation systems have become the main entry for users to discover products.
Meanwhile, with increasing concerns on the transparency and accountability of AI systems, studies on explainable AI have received more attention in both academic communities and industry~\cite{gilpin2018explaining,du2018techniques}.
Specifically in the domain of e-commerce information retrieval, explainability means the ability of a product retrieval system in providing explanations that allow users to understand, trust, and effectively control the retrieved products.
Previous studies have shown that providing recommendation results together with explanations on why the items are retrieved not only increases the conversion rates from clicks to purchases, but also improves user's satisfaction on e-shopping websites~\cite{zhang2014explicit}.
Thus, how to improve the explanability of product retrieval systems has become an important challenge and opportunity for e-commerce.


Interestingly, despite of the extensive studies on explainable product recommendation~\cite{herlocker2000explaining,bilgic2005explaining,tintarev2007effective,mcauley2013hidden}, the effectiveness and the potentials of explainable product search is mostly unexplored. 
As of today, more than 80\% of shoppers find products starting from search online\footnote{\url{https://www.retaildive.com/news/87-of-shoppers-now-begin-product-searches-online/530139/}}, which means that product search is still the most popular method to find products on e-commerce platforms.
On the one hand, developing product search engines is similar to developing product recommendation systems from multiple perspectives, including the need of personalization~\cite{ai2017learning}, the model of heterogeneous information~\cite{zamani2020learning}, etc.
On the other hand, by explicitly formulating and feeding a query to the systems, user's requirements and expectations for product search engines are significantly different from those for product recommendation.
For instance, while it is preferable to recommend PC games to a customer who recently purchased Alienware gaming laptops, it may not be a good idea when the user is searching for ``running shoes''.
Thus, how to retrieve and explain search results based on both the explicit need and implicit preferences of product search users make explainable product search an unique challenge in explainable information retrieval.


Existing studies on explainable AI can be broadly categorized into two directions, namely \textit{model-intrinsic} (or pre-hoc) interpretability and \textit{model-agnostic} (or post-hoc) interpretability~\cite{lipton2018mythos}. 
Model-intrinsic interpretability focuses on the construction of transparent AI systems that can explicitly explain its behavior based on its inference process.
In contrast, model-agnostic interpretability focuses on explaining model outputs without knowing the internal mechanism of the model.
Previous studies on explainable IR have explored both paradigms in document retrieval~\cite{singh2019exs,fernando2019study,singh2020model,10.1145/3331184.3331377} by creating pre-hoc or post-hoc explanations with text-matching signals extracted by the retrieval models from query-document pairs. 
In product search, however, it has been shown that text matching is relatively less important~\cite{10.1145/3336191.3371780,ai2019zero} comparing to other information such as knowledge entities and their relationships~\cite{guo2019attentive,liu2020structural} in determining user's purchase decisions.
Thus, how to create model-intrinsic/agnostic explanations in product search and how those two approaches would benefit or affect the development of explainable product search systems is mostly unknown.
To the best of our knowledge, the only study on explainable product search is the Dynamic Relation Embedding Model~\cite{ai2019explain} (DREM) that utilizes product knowledge graph to generate post-hoc result explanations.
For evaluation, however, Ai et al.~\cite{ai2019explain} simply use a survey to examine whether users are more likely to purchase after seeing the explanations and conduct no comparison of different explanation methodologies as well as possible factors that affect the effectiveness of search explanations. 

To fill in this blank, we propose to construct and train an intrinsic-explainable model for product search with user-interaction data and knowledge graph.
Inspired by the Zero Attention mechanism~\cite{ai2019zero}, we propose to extend DREM with a Hierarchical Gated Network (HGN) that explicitly construct user representations from items and knowledge entities related to the user's purchase history.
By extracting the attention weights from HGN, our proposed model is capable of generating model-intrinsic explanations for product search results.
To understand the advantages and drawbacks of pre-hoc and post-hoc search explanations, we conduct a crowdsourcing study with Amazon Mechanical Turk to evaluate and analyze the performance of model-agnostic and model-instrinsic explanations generated by the original DREM and the proposed DREM with HGN.
Experiment results show that model-intrinsic explanations usually could be more informative and reliable while model-agnostic explanations could have better potentials in attracting users to purchase the product.
Further, we propose an explanation performance task and build models to explore the possibility of automatically evaluating search explanations without human annotations.
Based on feature analysis, we find that the fidelity of search explanations could be more important for user's overall satisfaction with the search engines while the novelty of the search explanations could be more useful in attracting users to purchase the item. 

\section{Related Work}\label{sec:related_work}

There are three lines of studies that are important to our work: Interpretable AI, Explainable IR and Product Search.

\textbf{\textit{Interpretable AI}}.
The research of interpretable and explainable AI is a growing topic as the concerns on transparency and accountability of AI systems have increased dramatically recently~\cite{gilpin2018explaining}. 
In general, existing studies on interpretable AI can be broadly categorized into two groups, i.e., the studies on explaining machine learning (ML) models based on their internal structures, and the studies on explaining model outputs by treating the ML model as a black box~\cite{lipton2018mythos}.
Examples of the first group including the examination of network neurons and layers~\cite{nguyen2016synthesizing,bau2017network,frankle2018lottery,yosinski2014transferable,sharif2014cnn}, the use of attention networks~\cite{vaswani2017attention,wiegreffe2019attention,jain2019attention}, and the design of disentangled model structure and information representations~\cite{cramer2008effects,burgess2017understanding,higgins2017beta,locatello2018challenging}.
Examples of the second group including the construction of proxy models with linear classifiers~\cite{ribeiro2016should}, decision trees~\cite{schmitz1999ann,zilke2016deepred}, extracted rules~\cite{andrews1995survey,fu1994rule}, and salience map~\cite{simonyan2013deep,zeiler2014visualizing}.
Both paradigms have their own advantages and disadvantages depending on application scenarios, 
and the field of interpretable AI is still young with numerous new studies and approaches emerging every year~\cite{gilpin2018explaining}.

\textbf{\textit{Explainable Recommendation}}.
The studies of explainable retrieval systems have drawn the attention of researchers mainly starting from the last decade. 
Early IR systems based on term matching are transparent and explainable in nature~\cite{ponte1998language,robertson2009probabilistic,tintarev2015explaining}.
However, as more state-of-the-art retrieval systems rely on complex ML and latent representation models~\cite{mitra2018introduction,guo2019deep}, interpretability is no longer a minor problem for IR.
Most existing studies on explainable IR focus on recommendation tasks~\cite{zhang2020explainable}.
For example, model-based explainable recommendation methods attempt to develop models that generate both recommendations and explanations together~\cite{tintarev2007survey,zhang2016explainable,burke2002hybrid}.
Peake and Wang~\cite{peake2018explanation} created post-hoc explanations based on the latent vectors in recommendation models;
Zhang et al.~\cite{zhang2014explicit} explained recommendation results with facets extracted from user reviews.
Another line of explainable recommendation research focuses on analyzing the nature of user behaviors to help users better understand recommendations~\cite{herlocker2000explaining, herlocker2000understanding,balog2020measuring}.
Bilgic and Mooney~\cite{bilgic2005explaining} used statistical histrograms as explanations to help users understand rating distribution; Tintarew and Mashthoff~\cite{tintarev2007effective} provided user-centered design approaches to analyze the explanation effectiveness. 

\textbf{\textit{Explainable Search}}.
Search is fundamentally different from recommendation as user intents are explicitly expressed with queries.
Different from explainable recommendation, the studies on explainable search mostly focuses on the domain of ad-hoc retrieval, i.e., retrieving text documents such as news articles or web pages based on user's query.
For example,  Zeon Trevor et al.~\cite{fernando2019study} proposes to use DeepSHAP~\cite{lundberg2017unified} to explain the outputs of neural retrieval models; Verma and Ganguly~\cite{verma2019lirme} explore different sampling methods to build explanation models for a given retrieval model and proposes a couple of metrics to evaluate the explanations based on the terms in queries and documents.
Unfortunately, those methods are not applicable to product search as they are purely designed for text retrieval and text matching signals are relatively unimportant~\cite{10.1145/3336191.3371780,ai2019zero} compared to other information such as entity relationships and user purchase history in determining user's purchases.
As for how to create result explanations with heterogeneous entity and information in product search, to the best of our knowledge, the only study on this topic is proposed by Ai et al.~\cite{ai2019explain} that construct a dynamic relation embedding model to incorporate product knowledge graph and use it to explain product search results.
However, they only conducted a laboratory study to examine the effectiveness of the explanations generated by their model and did no comparison and study on different explanation methodologies as well as what factors are important for product search explanations.


\textbf{\textit{Product Search}}.
Early studies on product search focus on retrieving products based on structured product facets such as brands and categories~\cite{lim2013semantic,duan2013probabilistic,duan2013supporting}.
However, as there exists a significant vocabulary gap between user queries and product descriptions~\cite{van2013deep, nurmi2008product}, state-of-the-art approaches usually conduct product search in latent space with deep learning techniques~\cite{guo2018multi, wang2020metasearch,bi2019study}.
For example, Bi et al.~\cite{bi2019negative,10.1145/3404835.3462911} extract fine-grained review information with embedding networks; 
Guo et al.~\cite{guo2019attentive} model long/short term user preferences with attention networks over user query history.
There are also considerable studies on extracting ranking features and applying learning-to-rank methods for product search~\cite{aryafar2017ensemble,hu2018reinforcement,karmaker2017application,wu2017ensemble,carmel2020multi}.
In this paper, our main focus is not to build the state-of-the-art product search models but to explore how to build effective search explanations to better improve user experience.



\section{Methodology}\label{sec:model}

In this section, we describe our proposed method for explainable product search.
We start from introducing the framework of latent product retrieval models, the structure of the state-of-the-art explainable product search model (i.e., DREM) for model-agnostic explanation, and then propose a hierarchical gated network (HGN) to extend DREM for model-intrinsic search explanations. 

\subsection{Latent Product Retrieval Framework}

As discussed previously, the goal of product search is to retrieve products according to user's needs so that we can maximize the average transaction rate (i.e., user purchases) in search sessions.
Usually, this means ranking and showing products to users according to their probabilities to be purchased~\cite{van2016learning,ai2017learning}. 
Different from traditional IR tasks such as ad-hoc retrieval, information in product search is often stored in heterogeneous forms and classic retrieval models based on text matching often performs suboptimal in practice~\cite{nurmi2008product,guo2018multi,ai2019explain}.
Therefore, the state-of-the-art methods in product search often build retrieval models in latent spaces by representing and matching queries, users, and items with latent vectors.

In general, user's purchase decisions are affected by two factors~\cite{ai2017learning,guo2018multi,ai2019explain}: (1) the explicit purchase intents in the current session, which are usually expressed by user's queries, and (2) the implicit preferences over product properties (e.g., colors and brands), which are usually inferred from user's historical behaviors (e.g., previous purchases). 
Formally, let $\bm{q}$, $\bm{u}$, $\bm{i}\in\mathbb{R}^{\alpha}$ be the $\alpha$ dimensional latent representations of the search query, the user's personal preferences, and the item, respectively. 
Following previous studies~\cite{ai2017learning,guo2019attentive}, we model the probability of an item $i$ being purchased by a user $u$ after submitting a query $q$ with a latent generative model as
\begin{equation}
P(i|u,q) = \frac{\exp (\bm{i} \cdot \bm{S_{uq}})}{\sum_{i'\in I}\exp (\bm{i}' \cdot \bm{S_{uq}})}, ~~\bm{S_{uq}} = \bm{q} + \bm{u}
\label{equ:P_iuq}
\end{equation} 
where $I$ is the universal set of candidate items, and $S_{uq}$ is the latent representation of the user's purchase intent in search, which could be modeled as the linear combination of $\bm{q}$ and $\bm{u}$.\footnote{For simplicity, we ignore the discussions of more complicated models for $S_{uq}$ as it is not the focus of this paper.}

Under this formulation, the representations of queries, users, and items can be directly optimized for product search by maximizing the log likelihood of observed user purchases in search defined as
\begin{equation}
	\mathcal{L} =\log \!\prod_{u,q,i}\!P(i|u,q)=\!\!\sum_{u,q,i}\!\! \big(\bm{i}\! \cdot \!(\bm{q} \!+\! \bm{u}) \!-\! \log\! \sum_{i'\in I}\!\!\exp (\bm{i}' \!\cdot\! (\bm{q} \!+\! \bm{u}))\big)
	\label{equ:final_log_likelihood}
\end{equation}
While directly computing $\mathcal{L}$ is prohibitive due to the softmax function and the large number of items in $I$, there are many effective and mature solutions built with approximation algorithms such as hierarchical softmax and negative sampling~\cite{mikolov2013efficient}.
In this paper, we adopt the negative sampling strategy that approximate the denominator of softmax function by randomly sampling negative samples from $I$.
Therefore, the key problem of product search in latent space is how to construct the representations of queries, users, and items.

\subsection{Dynamic Relation Embedding Model and Model-agnostic Explanations}

To the best of our knowledge, the first model proposed for explainable product search is the Dynamic Relation Embedding Model (DREM)~\cite{ai2019explain}.
In order to utilize heterogeneous data and knowledge for product search and explanations, Ai et al.~\cite{ai2019explain} proposed to build a latent dynamic knowledge graph that jointly encodes the relationships between queries, users, items, as well as product-related knowledge entities. 
Specifically, the construction of DREM and search explanations include two parts: the modeling of entity relationships, and the extraction of explainable knowledge path between users and retrieved items.

\subsubsection{Product Knowledge Graph and Query Modeling}\label{sec:DREM_model}

Product search is different from product recommendation as the relevance and relationships between users and items could vary based on user's information need expressed in the search query.
To model both the static relationships between knowledge entities and the dynamic relationships between users, queries, and items, Ai et al.~\cite{ai2019explain} propose to adopt the TransE models~\cite{bordes2013translating} for product search and treat \textit{Search\&Purchase} as a special relationship that translates users to items. 
Formally, let $(h,r,t) \in \mathcal{G}$ be a relation triple with head entity $h$, relation $r$, and tail entity $t$ (e.g., \textit{IPhone} is \textit{Produced\_by} \textit{Apple}) in observed data $\mathcal{G}$.
Then DREM defines a linear translation function and a latent generative model to model $(h,r,t)$ as
\begin{equation}
	P(t|h,r) = \frac{\exp (\bm{t} \cdot (\bm{h} + \bm{r}))}{\sum_{t'\in T}\exp (\bm{t}' \cdot \bm{h} + \bm{r})}
	\label{equ:transE}
\end{equation}
where $T$ is the universal set of possible tail entity $t$, and $\bm{h}$, $\bm{r}$, $\bm{t}\in\mathbb{R}^{\alpha}$ are the embedding representations of the head entity, the relation, and the tail entity, respectively.
In other words, \textit{the entity $h$ can be translated to entity $t$ through relation $r$ with probability $P(t|h,r)$}.

As the relationship between users and items (i.e., \textit{Search\&Purchase}) varies according to different search queries, Ai et al.~\cite{ai2019explain} propose to create a dynamic relation embedding by encoding query string with a non-linear project function $\phi$ as
\begin{equation}
	\bm{q} = \phi(\{w_q|w_q \in q\}) = \tanh(W \cdot \frac{\sum_{w_q \in q}\bm{w_q}}{|q|} + b)
	\label{equ:f}
\end{equation}
where $w_q$ and $\bm{w_q}\in\mathbb{R}^{\alpha}$ are query words and their corresponding embedding representations, $W\in\mathbb{R}^{\alpha \times \alpha}$ and $b\in\mathbb{R}^{\alpha}$ are model parameters, and $\bm{q}\in\mathbb{R}^{\alpha}$ is the query embedding as well as the relation embeddding of \textit{Search\&Purchase}.
The probability of a user $u$ searched and purchased an item $i$ is then computed in the same way with other relation triples as shown in Eq.~(\ref{equ:transE}).

To optimize the embedding representations of all entities and relations for product search, DREM directly maximizes the log likelihood of all observed relation triples as
\begin{equation}
	\begin{split}
		\mathcal{L} = \!\!\!\sum_{(u,q,i)}& \!\!\!\log P(i|u,q) +  \!\!\!\sum_{(h,r,t) \in \mathcal{G}} \!\!\! \log P(t|h,r)\\
		\approx \!\!\!\sum_{(u,q,i)}&\log\sigma\big((\bm{u}+\bm{q})\!\cdot \!\bm{i}\big) + k\!\cdot\! \mathbb{E}_{i'\sim P_i}[\log\sigma\big(\!\!-\!(\bm{u}+\bm{q})\!\cdot\! \bm{i'}\big)] \\
		+ \!\!\!\!\!\!\sum_{(h,r,t) \in \mathcal{G}}\!\!\!\!\!\!&\log\sigma\big((\bm{h}+\bm{r})\!\cdot \!\bm{t}\big) + k\!\cdot\! \mathbb{E}_{t'\sim P_t}[\log\sigma\big(\!\!-\!(\bm{h}+\bm{r})\!\cdot\! \bm{t'}\big)] \\
	\end{split}
	\label{equ:aggregated_loss}
\end{equation}
where $\sigma(x)$ is the sigmoid function (i.e., $\sigma(x)=\frac{1}{1+e^{-x}}$) and we apply a negative sampling strategy with sample size $k$. 
$P_i$ is defined as a uniform item noisy distribution and $P_t$ is defined as a frequency-based entity noisy distribution~\cite{van2016learning,ai2019explain}. 





\subsubsection{Post-hoc Search Explanations}

With the latent knowledge graph learned from observed product purchases and meta data, Ai et al.~\cite{ai2019explain} argues that DREM is capable of creating post-hoc search explanations for each item retrieved for a user-query pair.
Specifically, as all relations and entities are encoded in the latent space with the TransE models defined in Eq.~(\ref{equ:transE}), one can infer an arbitrary item from a user-query pair by finding a set of relations and intermediate entities that translate the joint representation of user and query (i.e., $\bm{S_uq}$) to the item representation (i.e., $\bm{i}$). 
Then, the path from the user to the item can be used to create an explanation of why the item is relevant to the user's search intent. 

Let $\{r_u^j\}$ (where $r_u^0$ is \textit{Search\&Purchase}) and $\{r_i^m\}$ be two sequences of relations that finally translate a user $u$ and an item $i$ to an entity in entity space $\Omega_e$.
DREM defines a soft matching path between $u$ and $i$ through $e\in \Omega_e$ with score: 
\begin{equation}
		\begin{split}
M(e | u, i) =& \log \big(P(e|u, \{r_u^j\})P(e|i, \{r_i^m\})\big) \\
 =& \log P(e|e_u) + \log P(e|e_i)\\
 =& \log\frac{\exp(\bm{e_u}\!\cdot\! \bm{e} - \gamma j)}{\sum_{e'\in \Omega_e}\!\!\exp(\bm{e_u}\!\cdot\! \bm{e'})} \!+\! \log\frac{\exp(\bm{e_i}\!\cdot\! \bm{e} - \gamma m)}{\sum_{e'\in \Omega_e}\!\!\exp(\bm{e_i}\!\cdot\! \bm{e'})} 
	\end{split} 
	\label{equ:soft_match}
\end{equation}
where $\bm{e_u}=\bm{u}+\sum_j \bm{r_u^j}$, $\bm{e_i}=\bm{i}+\sum_m \bm{r_i^m}$, and $\gamma$ is a hyper-parameter\footnote{Please refer to the original paper~\cite{ai2019explain} for more details.}. 
While the soft matching score of a path does not have any meanings to users, Ai et al.~\cite{ai2019explain} argues that it indicates the model's confidence on the path.
Thus, they sort all potential inference path from user-query pair to a target item with the soft matching score and directly create a search explanation using simple templates and the relations/entities on the path to explain why $i$ is retrieved for $u$ by $q$.  
For example, suppose that there is a path from user $u$ to Apple Pencil with query \textit{``tablet''} and relation \textit{Brought\_Together}, then we can explain why we retrieved Apple Pencil with a post-hoc explanation as ``Apple Pencil is retrieved because it is frequently \textit{Brought\_Together} with products retrieved by query ``tablet'' ''.




\subsection{Hierarchical Gated Network and Model-intrinsic Explanations}

While DREM can provide post-hoc explanations to search results with inference paths on knowledge graph, the retrieval process of the model is simply ranking items according to the dot product between user-query pair and item representation in the latent space, which are not necessarily correlated to the generated explanations.
In practice, we may prefer a transparent retrieval model that could provide direct explanations to its inference process for many reasons such as model reliability and result accountability~\cite{lipton2018mythos}.
Inspired by the Zero Attention Mechanism (ZAM)~\cite{ai2019zero}, in this paper, we propose an extension to DREM to enhance it's interpretability and enable it to provide model-intrinsic search explanations.

\subsubsection{Attention Network with Gates}

ZAM is first proposed to conduct selective personalization in product search~\cite{ai2019zero}.
The idea of ZAM is to relax the assumption of traditional attention mechanism by allowing the network to attend none input data when the query is not relevant to any input vectors.
Let $\bm{q}$ be the query vector and $\bm{X}$ be the input vectors of an attention network, then ZAM computes the output $\bm{y}$ by attending $\bm{q}$ to both $\bm{X}$ and a zero vector $\bm{0}$ as 
\begin{equation}
	\bm{y} = \sum_{x \in X}\frac{\exp(f(\bm{q},\bm{x}))}{\exp(f(\bm{q}, \bm{0})) + \sum_{x' \in X}\exp(f(\bm{q},\bm{x'}))}\bm{x}
	\label{equ:ZAM}
\end{equation}   
where $\bm{0}$ is a vector with all elements equal to 0, and $f(\bm{q},\bm{x})$ is the attention function that computes the attention score of $x$ with $q$. 

By adding $\bm{0}$ to the attention network, ZAM naturally creates a gate that controls whether the output vector of the attention network would be fed into downstream applications or not. 
Let $\bm{a}_X$ be the vector of $\{f(\bm{q}|\bm{x})|x\in X\}$, then ZAM can be reformulated as
\begin{equation}
	\bm{y} = \frac{\exp(\bm{a}_X)}{\exp(f(\bm{q}, \bm{0})) + \exp^+(\bm{a}_X)}\cdot \bm{X}
	\label{equ:ZAM_sig}
\end{equation}
where $exp^+(\bm{a}_X)$ is the element-wise sum of $exp(\bm{a}_X)$.
Thus, the output $\bm{y}$ would be influenced by the input $X$ only when the aggregated attention of $X$ is significantly larger than a threshold $f(\bm{q}, \bm{0})$.

\begin{figure*}
	\centering
	\includegraphics[width=4.8in]{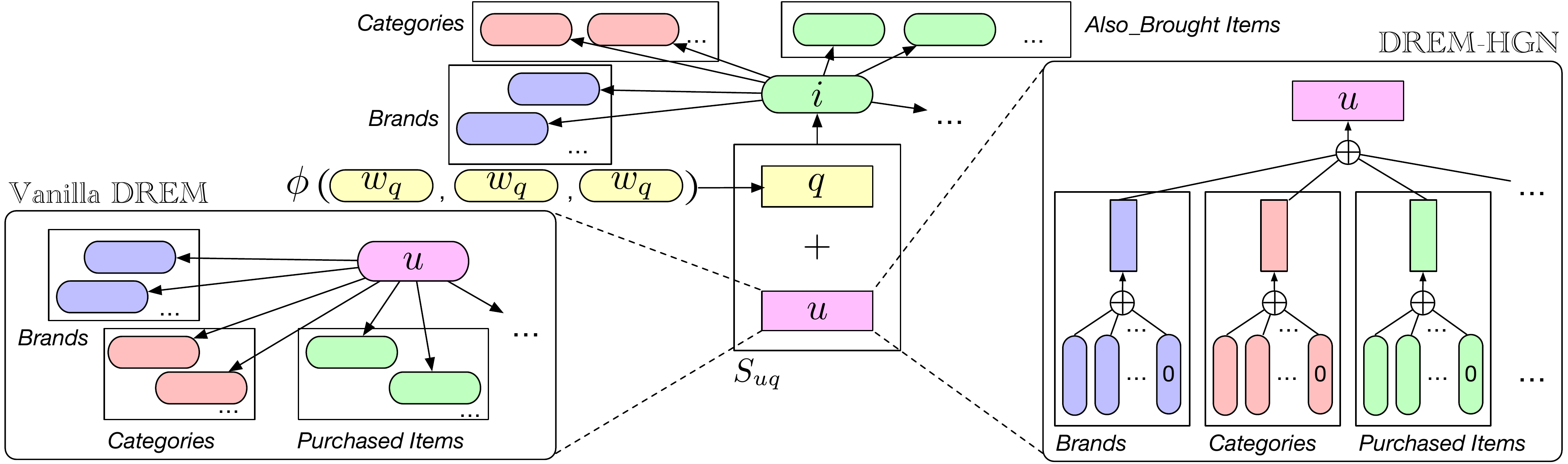}
	\caption{An illustration of the vanilla DREM and DREM with HGN. Different types of entities are colored differently. Squashed rectangles are vectors randomly initialized and learned in training, and rectangles are vectors encoded from other vectors.}
	\vspace{-12pt}
	\label{fig:model}
\end{figure*}

\subsubsection{User Modeling in Hierarchy}
We now describe how we extend DREM to a transparent product search model with the idea of ZAM. 
The construction of a interpretable retrieval model with heterogeneous product knowledge involves two questions: (1) how to model user preferences in a specific knowledge domain according to the current search query, and (2) how to jointly combine user preferences in each knowledge domain to retrieve items for the current search query.
To solve these questions, we proposed to build a Hierarchical Gated Network (HGN) to model user preferences in search with knowledge entities associated to user's purchase history.
A illustration of the DREM with HGN is shown in Figure~\ref{fig:model}.

Formally, let $\Omega_e^u$ be the set of entities with type $e$ associated to a user $u$. 
For example, $\Omega_e^u$ could be the items or brands purchased by $u$. 
For each knowledge domain $e$, we compute a latent embedding $\bm{u}_e$ for $u$ in $e$ by attending each entity with the current query $q$ as
\begin{equation}
	\bm{u}_e = \sum_{e \in \Omega_e^u}\frac{\exp(f_e(\bm{q},\bm{e}))}{\exp(f_e(\bm{q}, \bm{0})) + \sum_{e' \in \Omega_e^u}\exp(f_e(\bm{q},\bm{e'}))}\bm{e}
	\label{equ:ZAM_e}
\end{equation}  
where $f_e(\bm{q},\bm{e})$ is a simple attention function defined as
\begin{equation}
	f(\bm{q}, \bm{e}) = \big(\bm{e}\cdot \tanh(\bm{W}_e^f \cdot \bm{q} + \bm{b}_{e})\big) \cdot \bm{W}_e^h
	\label{equ:attention_function}
\end{equation}
where $\bm{W}_e^h\in \mathbb{R}^{\beta}$, $\bm{W}_e^{f} \in \mathbb{R}^{\alpha \times \beta \times \alpha}$, $\bm{b}_{e} \in \mathbb{R}^{\alpha \times \beta}$, and $\beta$ is a hyper-parameter that controls the number of the attention heads. 

To further aggregate $\bm{u}_e$ from each domain to create the final user embedding $\bm{u}$, we apply another layer of zero attention above all knowledge domains.
Let $\Omega^u = \{\Omega_e^u\}$, then 
\begin{equation}
	\bm{u} = \sum_{\Omega_e^u \in \Omega^u}\frac{\exp(f_u(\bm{q},\bm{u}_e))}{\exp(f_u(\bm{q}, \bm{0})) + \sum_{\Omega_{e'}^u \in \Omega^u}\exp(f_u(\bm{q},\bm{u}_e'))}\bm{u}_e
	\label{equ:HGN}
\end{equation} 
where $f_u(\bm{q},\bm{u}_e))$ is another attention function with similar form of Eq.~(\ref{equ:attention_function}) but different set of parameters.

Intuitively, the idea of HGN is to construct a hierarchical zero attention network that aggregates fine-grained user preferences from each knowledge domain to a unified user vector based on the search query. 
For parameter optimization, we simply follow the methodology of DREM introduced in Section~\ref{sec:DREM_model} and replace the original user vector $\bm{u}$ with the new user vector constructed by HGN.
Through this way, we can easily track down the usage of each knowledge entity in product search and create a higher-level of transparency and interpretability to DREM.


\subsubsection{Pre-hoc Search Explanations}

The advantage of HGN-based DREM is its ability to create model-intrinsic explanations. 
Attention network is explainable by nature as the importance of input data is directly reflected by their attention weights in model outputs.
With the help of HGN, we can not only infer the importance of each user-associated entity in building the final retrieval model (i.e., $S_{uq}$ in Eq.~(\ref{equ:P_iuq})), but also distinguish how much utility is obtained from understanding user's preferences over retrieved items or the general relevance/popularity between items and search queries.

For pre-hoc search explanations, the attention weights in HGN can be split into two parts.
The first part is the attention score of each user-associated knowledge domain and entity. 
Let $A^e_{\Omega_e^u}$ be the attention weight that entity $e$ received within domain $\Omega_e^u$, and $A^u_{\Omega_e^u}$ be the attention weight that domain $\Omega_e^u$ received in search, then
\begin{equation}
	\begin{split}
		A^e_{\Omega_e^u} =& \frac{\exp(f_e(\bm{q},\bm{e}))}{\exp(f_e(\bm{q}, \bm{0})) + \sum_{e' \in \Omega_e^u}\exp(f_e(\bm{q},\bm{e'}))} \\
		A^u_{\Omega_e^u}=& \frac{\exp(f_u(\bm{q},\bm{u}_e))}{\exp(f_u(\bm{q}, \bm{0})) + \sum_{\Omega_{e'}^u \in \Omega^u}\exp(f_u(\bm{q},\bm{u}_e'))}
	\end{split}
\label{equ:HGN_entity_attention}
\end{equation}
Intuitively, $A^u_{\Omega_e^u}$ is the importance of domain $\Omega_e^u$ in building the final user model $u$, and $A^e_{\Omega_e^u}$ is the importance of entity $e$ in the domain. 
To explain the behavior of the product search model with these information, we can adopt simple templates to generate user readable search explanations with the attention weights.
For example, for a specific user-query pair, if the attention weight of the domain \textit{Brand} is 0.5, and \textit{Apple} is the entity that received the highest attention within \textit{Brand}, we can generate a pre-hoc search explanations as ``this product are retrieved 50\% because of the \textit{Brand} of products previously purchased by the user, such as \textit{Apple}''.

The second part of the attention weights in HGN is the attention on the zero vector. 
As depicted in Figure~\ref{fig:model}, HGN allows the model to attend to a zero vector when aggregating the information extracted from each knowledge domain with weight $A_0^u$ as
 \begin{equation}
 	\begin{split}
 		A^u_{0} =& \frac{\exp(f_u(\bm{q},\bm{0}))}{\exp(f_u(\bm{q}, \bm{0})) + \sum_{\Omega_{e'}^u \in \Omega^u}\exp(f_u(\bm{q},\bm{u}_e'))}
 	\end{split}
 	\label{equ:HGN_query_attention}
 \end{equation}
Particularly, we apply a negative sampling strategy to maximize the probability of observed purchase $P(i|u,q)$, which has been proven to be equivalent to factorizing the pointwise mutual information between $\bm{S_{uq}}$ and $\bm{i}$~\cite{levy2014neural}. 
According to Eq.~(\ref{equ:P_iuq}) and (\ref{equ:HGN_query_attention}), when $A^u_{0}$ is close to 1, $\bm{S_{uq}}$ would downgrade to $\bm{q}$ and the final retrieval model is essentially retrieving items according to their mutual information with the query. 
From this perspective, the weight of the zero vector in HGN can be seen an indicator of the importance of item popularity under the query in the generation of the final ranked list.
Therefore, given a particular $A_0^u$, we could explain the results retrieved by HGN as ``this product is retrieved $A_0^u$\% because of its popularity under the query''.








\section{Retrieval Experiments}\label{sec:setup}

In general, the evaluation of an explainable product search model involves two parts: (1) the evaluation of retrieval performance in terms of retrieving items that are most likely to be purchased by users, and (2) the evaluation of explanation effectiveness in terms of illustrating the connections between users, queries, and retrieved items as well as increasing the conversion rates from search to purchase. 
In this section, we focus on the first part and introduce our settings and results in retrieval experiments.


\subsection{Experimental Setup}

The goal of retrieval experiments is to evaluate the effectiveness of product search models in retrieving relevant items for user-query pairs. 
To this end, we conduct experiments on a well-established product search dataset and implement a couple of state-of-the-art baselines to analyze the performance of DREM with HGN.

\subsubsection{Dataset}

Our testbed is a well-established Amazon product search benchmark datasets~\cite{van2016learning,ai2017learning,guo2019attentive}.
The dataset contains user's purchases, reviews, and queries in a variety of categories as well as detained descriptions and meta data of a large number of items on Amazon\footnote{Please refer to~\cite{van2016learning,ai2017learning,guo2019attentive} for the details of Amazon search datasets.}. 
Specifically, we conduct experiments on three categories, i.e., \textit{Electronics}, \textit{Health\&PersonalCare}, and \textit{Sports\&Outdoors}, and use the 5-core data where each user/item has at least 5 reviews~\cite{ai2017learning,guo2019attentive}.

\begin{table}
	\centering
	\caption{Statistics for the 5-core data.}
	\scalebox{0.68}{
		\begin{tabular}{l  c  c  c } 
			\toprule
			& \textit{Electronics}& \textit{Health\&PersonalCare} & \textit{Sports\&Outdoors}\\
			\midrule
			
			Vocabulary size & 142,922 & 38,772 & 32,386\\ %
			Number of reviews & 1,689,188 & 346,355 & 296,337 \\ %
			Number of users & 192,403 & 38,609 & 35,598\\ %
			Number of items & 63,001 & 18,534 & 18,357\\ %
			Number of brands & 3,525 & 3,855 & 2,412\\ %
			Number of categories & 983 & 861 & 1,443\\ %
			
			\midrule
			
			\textit{Also\_bought} per item & 36.70$_{\pm 38.56}$ & 63.03$_{\pm 35.36}$ & 75.18$_{\pm 31.98}$\\ %
			\textit{Also\_viewed} per item & 4.36$_{\pm 9.44}$ & 15.43$_{\pm 9.35}$ & 14.46$_{\pm 12.24}$ \\ %
			\textit{Bought\_together} per item & 0.59$_{\pm 0.72}$ & 0.86$_{\pm 0.77}$ & 0.83$_{\pm 0.76}$\\ %
			\textit{Brand} per item & 0.47$_{\pm 0.50}$ & 0.76$_{\pm 0.43}$ & 0.67$_{\pm 0.47}$ \\ %
			\textit{Category} per item & 4.39$_{\pm 0.95}$ & 4.20$_{\pm 0.93}$ & 4.82$_{\pm 1.33}$\\ %
			
			\midrule
			
			Number of reviews & 1,275,432/413,756 & 261,281/85,074 & 224,807/71,530\\ %
			Number of user-query pairs & 1,204,928/5,505 & 232,187/207 & 214,919/1,739\\ 
			Relevant items per pair & 1.12$_{\pm 0.48}$/1.01$_{\pm 0.09}$ & 1.13$_{\pm 0.47}$/1.00$_{\pm 0.00}$ & 1.12$_{\pm 0.45}$/1.01$_{\pm 0.13}$\\ 
			\bottomrule
		\end{tabular}
	}
	\label{tab:dataset_statistics}
\end{table}

Other than review text, to incorporate rich product meta data for product search, we also consider five types of entity relationships in our experiments.
They are \textit{Also\_bought}: users who purchased item $i_1$ has also purchased item $i_2$ ($i_1\rightarrow i_2$); \textit{Also\_viewed}: users who viewed item $i_1$ also viewed item $i_2$ ($i_1\rightarrow i_2$); \textit{Bought\_together}: item $i_1$ was purchased together with item $i_2$ in a single transaction ($i_1\rightarrow i_2$); \textit{Brand}: item $i$ has brand $b$ ($i\rightarrow b$); and \textit{Category}: item $i$ has category $c$ ($i\rightarrow c$). 
More data statistics can be found in Table~\ref{tab:dataset_statistics}.

\subsubsection{Baselines}

Other than the naive DREM proposed by Ai et al.~\cite{ai2019explain}, in our experiments, we include six state-of-the-art product search baselines including classic retrieval models such as
\begin{itemize}
	\item \textbf{QL}: the query-likelihood model~\cite{ponte1998language} that ranks items according to the log likelihood of queries in the unigram language model built with item descriptions and reviews.
	\item \textbf{BM25}: the classic probabilistic model proposed by Robertson and Walker~\cite{robertson2009probabilistic} built on the item's descriptions and reviews.
	\item \textbf{LTR}\footnote{We extract ranking features for LTR following the same method used by Ai et al.~\cite{ai2019explain}, which is ignored in this paper due to page limit.}: a learning-to-rank model built with LambdaMART.
\end{itemize}
and latent product search baselines such as 
\begin{itemize}
	\item \textbf{LSE}: the Latent Semantic Entity model~\cite{van2016learning} that ranks items based on the similarity of queres and items in latent spaces.
	\item \textbf{HEM}: the Hierarchical Embedding Model~\cite{ai2017learning} that personalizes product search results with a latent retrieval framework.
	\item \textbf{ZAM}: the original Zero Attention Model~\cite{ai2019zero} that conducts selective personalization in product search.
\end{itemize}

\subsubsection{Implementation and Evaluation Details}

Following previous studies~\cite{guo2019attentive,ai2019explain}, we partition the data in each product category by randomly hiding 30\% user purchases from the training process and use them as the testing data.
We randomly select 30\% queries as the test queries and match users with queries extracted from their purchase history to form training and testing user-query pairs. 
A item is considered relevant to a user-query pair when it is relevant to the query and has been purchased by the user. 
More information about our data partition can be found in Table~\ref{tab:dataset_statistics}.

For implementation details, we follow the settings proposed by Ai et al.~\cite{ai2019explain} by building QL and BM25 with galago\footnote{https://sourceforge.net/p/lemur/wiki/Galago/}, building LTR with ranklib\footnote{https://sourceforge.net/p/lemur/wiki/RankLib/}, and tuning the Dirichlet smoothing parameter $\mu$ in QL from 1000 to 3000, the scoring parameter $k$ and $b$ in BM25 from 0.5 to 4 and 0.25 to 1, respectively.
The number of trees and leaf in LambdaMART model used in LTR are set as 1000 and 10, and we tune the learning rate from 0.01 to 0.1.
For latent product retrieval models such as LSE, HEM, ZAM, the vanilla DREM~\cite{ai2019explain} and our extended DREM with HGN (DREM-HGN), we use Adagrad~\cite{luo2019adaptive} with batch size 64 to optimize the latent vectors and set the sample size of negative sampling as 5.
We clipped batch gradients with norm 5 to avoid unstable updates and train each model for 20 epochs by gradually decrease the learning rate from 0.5 to 0 (note that most models converge after 10 epoches).
For fair comparison, we fixed the personalization weight $\eta$ in HEM, ZAM, DREM, and DREM-HGN as 0.5 (which results in Eq.~(\ref{equ:P_iuq})) and the size of all latent vectors as 100 (i.e., $\alpha=100$). 
We acknowledge that having larger vector size could boost the performance of some latent product search models, especially those using rich product knowledge and information (e.g., DREM and DREM-HGN)~\cite{ai2019explain}.
However, this is the not focus of this paper and we ignore the tunning of $\alpha$ so that the embeddings learned by different models have comparable dimentionalities.

We adopt mean average precision (MAP), mean reciprocal rank (MRR) and normalized discounted cumulative gain (NDCG) to evaluate the performance of product search models. 
For each user-query pair, we retrieve 100 items among all candidate items in each dataset to generate the rank list and compute MAP and MRR accordingly.
We also report NDCG with cutoff 10 and 50.
Significant tests are measured by the Fisher randomization test~\cite{smucker2007comparison} with p < 0.05.




\subsection{Retrieval Results}

\begin{table*}[t]
	\centering
	\small
	\caption{The retrieval performance of product search models. Best performance in each model group are highlighted in bold. $*$ and $\dagger$ denote significant improvements over the best classic retrieval baselines and latent search baselines, respectively.}
	\scalebox{0.85}{
	\begin{tabular}{ c || c | c | c | c || c | c | c | c || c | c | c | c   } 
		\hline
		& \multicolumn{4}{c||}{\textit{Electronics}} & \multicolumn{4}{c||}{\textit{Health\&PersonalCare}} & \multicolumn{4}{c}{\textit{Sports\&Outdoors}}\\ \hline 
		Model  & MAP & MRR & NDCG@10 & NDCG@50 & MAP & MRR & NDCG@10 & NDCG@50 & MAP & MRR & NDCG@10 & NDCG@50  \\\hline
		\hline
		QL & 0.166 & 0.164 & 0.187 & 0.210 & 0.063 & 0.063 & 0.059 & 0.097 & 0.068 & 0.067 & 0.080 & 0.117  \\ \hline
		BM25 & 0.216 & 0.213 & 0.227 & 0.270 & \textbf{0.076} & \textbf{0.076} & \textbf{0.088} & \textbf{0.131} & 0.092 & 0.091 & 0.097 & 0.142 \\ \hline
		LTR & \textbf{0.216} & \textbf{0.216} & \textbf{0.230} & 0.303 & 0.060 & 0.060 & 0.055 & 0.113 & \textbf{0.109}$^\dagger$ & \textbf{0.109}$^\dagger$ & \textbf{0.120}$^\dagger$ & \textbf{0.166} \\ \hline \hline
		LSE & 0.108 & 0.108 & 0.137 & 0.183  & 0.016 & 0.016 & 0.000 & 0.113 & 0.015 & 0.015 & 0.019 & 0.040\\ \hline
		HEM & 0.156 & 0.156 & 0.182 & 0.197 & 0.157$^*$ & 0.157$^*$ & 0.146$^*$ & 0.200$^*$ & 0.075 & 0.075 & 0.086 & 0.128\\ \hline
		ZAM & 0.115 & 0.115 & 0.130 & 0.162 & 0.208$^*$ & 0.208$^*$ & 0.244$^*$ & 0.263$^*$ & 0.074 & 0.075 & 0.087 & 0.153\\ \hline
		Vanilla DREM & \textbf{0.231}$^*$ & \textbf{0.232}$^*$ & \textbf{0.268}$^*$ & \textbf{0.314}$^*$ & \textbf{0.349}$^*$ & \textbf{0.349}$^*$ & \textbf{0.378}$^*$ & \textbf{0.429}$^*$ & \textbf{0.099} & \textbf{0.099} & \textbf{0.113} & \textbf{0.180}$^*$\\ \hline \hline
		DREM-HGN & \textbf{0.244}$^{*\dagger}$ & \textbf{0.245}$^{*\dagger}$ & \textbf{0.275}$^{*\dagger}$ & \textbf{0.339}$^{*\dagger}$ & \textbf{0.536}$^{*\dagger}$ & \textbf{0.536}$^{*\dagger}$ & \textbf{0.556}$^{*\dagger}$ & \textbf{0.588}$^{*\dagger}$ & \textbf{0.126}$^{*\dagger}$ & \textbf{0.127}$^{*\dagger}$ & \textbf{0.141}$^{*\dagger}$ & \textbf{0.215}$^{*\dagger}$\\ \hline \hline
	\end{tabular}
}
	\label{tab:retrieval_results}
	\vspace{-10pt}
\end{table*}

The results of our retrieval experiments are shown in Table~\ref{tab:retrieval_results}.
Similar to previous studies~\cite{van2016learning,ai2017learning}, we observed that latent product search models usually perform better than classic retrieval models constructed based on text matching signals. 
For example, our best latent product search baseline (i.e., the vanilla DREM) has significantly outperformed QL and BM25 on all the datasets.
This is reasonable as previous studies have observed significant vocabulary gap between queries and item descriptions~\cite{nurmi2008product,van2016learning}, and users often purchase items that ``seems'' irrelevant to their submitted query in text~\cite{10.1145/3336191.3371780}. 
After incorporating more complex behavior features such as item popularity, the LTR baseline has managed to outperform DREM on \textit{Sports\&Outdoors}, but still performs worse than DREM on \textit{Electronics} and \textit{Health\&PersonalCare}.

Among all latent product search models, the non-personalized baseline (i.e., LSE) performs the worst, which demonstrates the importance of personalization in product search.
In the results of personalized product search models, we observed that DREM has significantly outperformed other baselines with large improvements from 25\% to 50\%. 
This indicates that incorporating rich information from product knowledge graph and meta data is indeed helpful in improving the effectiveness of product search.
Further, our proposed model (i.e., DREM-HGN) has achieved the best performance in our experiments. 
It has outperformed all the baselines significantly and achieved 5.6\%, 53.6\%, and 28.2\% MRR improvements over the vanilla DREM on \textit{Electronics}, \textit{Health\&PersonalCare}, and \textit{Sports\&Outdoors}, respectively.
This indicates that user representations encoded from product knowledge with the hierarchical gated network is more useful than the user embedding learned by DREM from randomly initialed vectors.

While the results of DREM-HGN outperforming the vanilla DREM is not surprising as the former has incorporated a complicated attention network to model query-specific user preferences, the main advantage of HGN is its transparency and model-intrinsic interpretability that allows the generation of pre-hoc explanation for product search.
To compare the post-hoc and pre-hoc search explanations generated by DREM and DREM-HGN in detail, we further conduct a series of explanation evaluation and analysis. 



\section{Explanation Evaluation} \label{sec:exp_explanation}

In practice, both pre-hoc and post-hoc explanations have their unique advantages for IR.
For example, pre-hoc search explanations are considered more reliable as they are directly inferred from the structure of the retrieval model.
In contrast, post-hoc search explanations are more flexible as it neither enforces the retrieval model to have intrinsic interpretability nor requires access to the internal structure and data flow of the model.
To the best of our knowledge, DREM is the only explainable models for product search in the literature. 
Ai et al.~\cite{ai2019explain} has conducted a laboratory user study and show that the post-hoc search explanations extracted by DREM are useful in attracting users to purchase the items.
However, how this is achieved or what factors are important for the quality of search explanations are mostly unexplored.
In this section, we conduct experiments to evaluate and compare the pre-hoc explanations created by DREM-HGN with the post-hoc explanations created by the vanilla DREM for product search.
Specifically, we want to study and shed some lights on the following research questions:

\vspace{2pt}
\noindent\textbf{RQ1}: \textit{Which types of search explanations do users prefer? Model-intrinsic ones or model-agnostic ones?} 
\vspace{2pt}

\noindent\textbf{RQ2}: \textit{What factors are important for product search explanations?} 
\vspace{2pt}



\subsection{Experimental Setup}

We design and conduct a crowdsourcing experiment on Amazon Mechanical Turk\footnote{\url{https://requester.mturk.com/}} (AMT) to evaluate the pre-hoc and post-hoc search explanations generated by DREM-HGN and DREM.

\subsubsection{Explanation Generation}

As described in Section~\ref{sec:model}, both DREM and DREM-HGN create explanations with templates using knowledge entities and relations extracted by the models.
For fair comparison, we adopt a single set of explanation templates for both DREM and DREM-HGN. 
For example, given a specific relation triple such as (item, \textit{Brand}, Apple) extracted by the vanilla DREM, we would create an explanation as 
\textit{``This product was retrieved because the user often buys products with \textbf{brands} such as \textbf{Apple}''}.
\noindent Also, as DREM-HGN relies on the attention weights extracted from HGN to explain its behavior, we add the corresponding information in the template and create explanations such as
\textit{``This product was retrieved \textbf{50\%} because the user often buys products with \textbf{brands} such as \textbf{Apple}''}.

To avoid the randomness in search explanations and to improve the robustness of crowdsourcing, we increase the redundancy of our experiment by allowing each model to provide a group of explanations instead of a single one.
Specifically, we extracted and grouped the top-3 search explanations extracted by DREM and DREM-HGN and allow each explanation to include at most 3 relevant knowledge entities.
Instead of requiring each crowdsourcing worker to annotate each search explanation, we let workers to annotate the search explanations in groups so that the final results would be influenced less by the quality variance of explanations provided by each model.
We refer to the group of explanations generated by DREM and DREM-HGN as Model-Agnostic Explanation (MAE) and Model-Intrinsic Explanation (MIE), respectively.


\subsubsection{Annotation Strategy}

Previous studies~\cite{zhang2016explainable,wang2018explainable,ai2019explain} evaluated product recommendation and search explanations mainly from three perspectives: (1) whether the explanation has provided more relevant information about the item and the query, or \textit{Informativeness}; (2) whether the explanation is useful in attracting the user to purchase the item, or \textit{Usefulness}; and (3) whether providing the explanation would increase user's satisfaction for the service provided by the product search engine, or \textit{Satisfaction}.

In this paper, we adopt the same strategy to evaluate the performance of search explanations.
However, instead of requiring crowdsourcing workers to directly annotate each explanation with a 5-level score~\cite{wang2018explainable}, we propose to conduct pairwise comparisons for the explanations generated by DREM and DREM-HGN for each user-query-item triple and let the workers to annotate their pairwise preferences only.
Pairwise preferences have been proven to be much more robust and reliable comparing to pointwise relevance judgements in IR~\cite{joachims2017accurately}.
Through this way, we hope to improve the quality of our crowdsourcing experiments as well as exploring the possibility of building automatic search explanation evaluation models for product search, which is further discussed in Section~\ref{sec:classification}.



\subsubsection{Data Sampling}

Our crowdsourcing dataset is sampled from the retrieval experiment dataset of \textit{Electronics}.  
\textit{Electronics} is one of the most popular product categories on Amazon.
Products in \textit{Electronics} usually have less complicated knowledge structures (e.g., less entity relations per item as shown in Table~\ref{tab:dataset_statistics}) and are more familiar to workers on AMT.
Specifically, we randomly sampled 101 user-query pairs from the test data of \textit{Electronics} where both DREM and DREM-HGN achieved MRR scores greater or equal to 0.1. 
For fairness, we extracted the user-query-item triples to explain by pairing user-query pairs with the item purchased by the user in the corresponding session.
Thus, all sampled items are indeed purchased by the user and AMT workers only need to judge which explanations can better explain the user's purchase in the search session. 
Specifically, we recruited three workers per case, and applied a voting process to assign the final labels.


\begin{figure}
	\centering
	\includegraphics[width=3in]{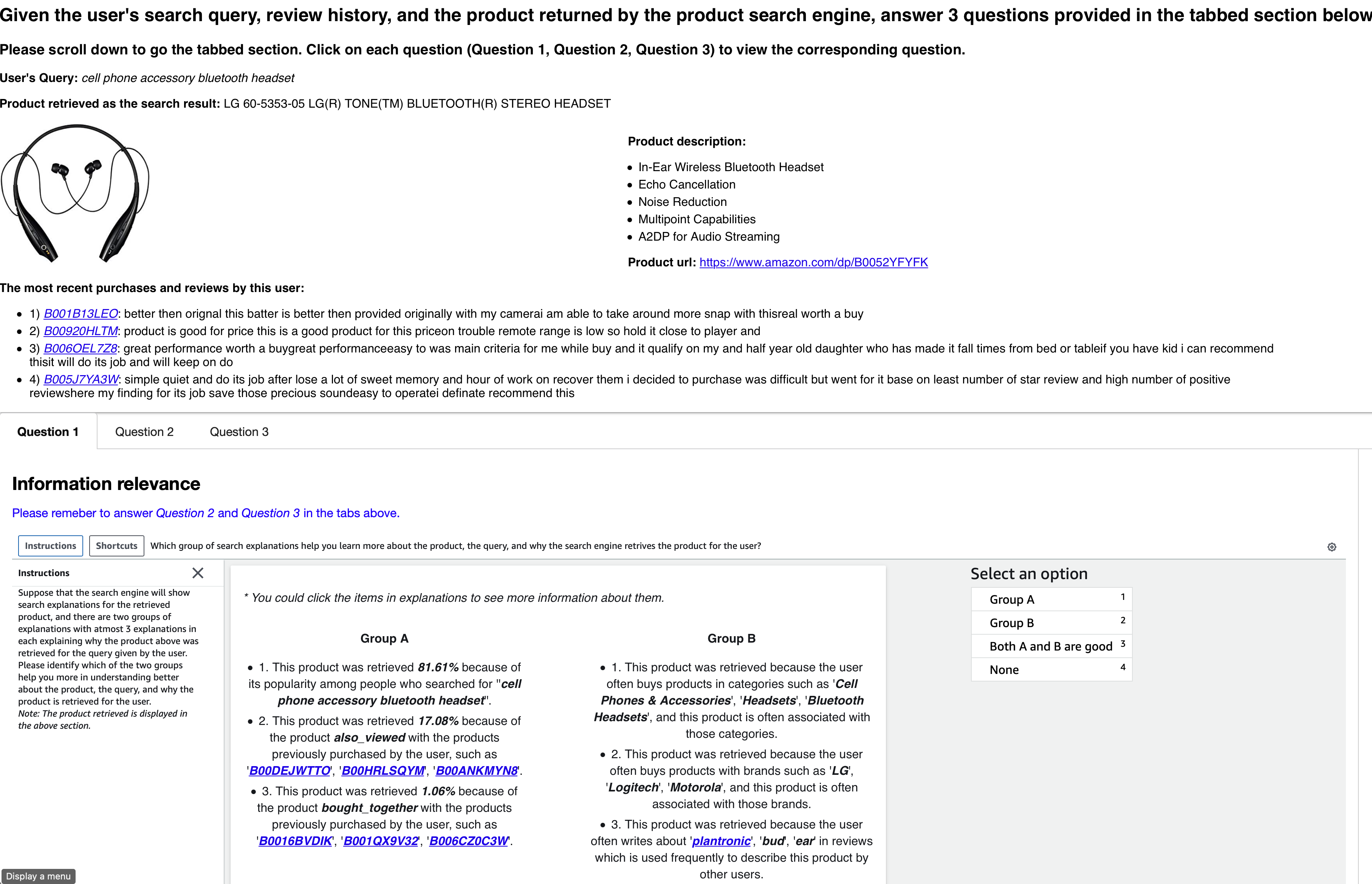}
	\caption{An illustration of the crowdsourcing UI.}
	\label{fig:UI}
\end{figure}

\subsubsection{UI Design}

Figure~\ref{fig:UI} provides an illustration of the UI we used for the crowdsourcing experiments.
On the top of the UI, we provided a variety of information related to the current item, including product links, images, titles, descriptions, the search query, and the recent purchases and reviews of the current user.
In the center of the UI, we implemented a tab-based frame that allows workers to navigate and annotate the informativeness, usefulness, and satisfaction of search explanations.
In each tab, we provided and asked workers to read a detailed instruction on the annotation process on the left, and show the groups of explanations created by DREM (i.e., MAE) and DREM-HGN (i.e., MIE) in the middle.
We also provided the original links on Amazon to all items and entities shown in the product descriptions, user reviews, or search explanations.
To avoid unnecessary biases in the annotation process, we anonymized MAE and MIE by randomly assigning them as ``Group A'' and ``Group B''.
Workers only need to click the buttons on the right to indicate which explanation group provide better search explanations: ``Group A'', ``Group B'', both, or none.
We eyeballed the collected data and manually filtered out workers with unreasonable behaviors.
The source code of our models, experiment platforms, and all crowdsourcing data can be found in links below\footnote{\url{https://github.com/utahIRlab/AMTurk-Product-Search-Explanation-Evaluation}}\footnote{\url{https://github.com/utahIRlab/drem-attention}}.




\vspace{-5pt}
\subsection{Crowdsourcing Results}

To answer \textbf{RQ1}, we show the results of our crowdsourcing experiment in Table~\ref{tab:crowd_results}.
As shown in table, most workers found that the model-intrinsic explanations provided by DREM-HGN (i.e., MIE) are preferable over the model-agnostic explanations provided by DREM (i.e., MAE) from the perspectives of \textit{Informativeness} and \textit{Satisfaction}.
This is not surprising as MIE provides more information about the actual inference process of the retrieval model  (e.g., the attention weights), which makes it more reliable and trustworthy to users.
However, in terms of \textit{Usefulness}, we do not observe any significant differences between MIE and MAE. 
The overall scores of MAE on \textit{Usefulness} is slightly higher than those for MIE. 
In fact, the Fleiss Kappa $\kappa$~\cite{viera2005understanding} of binary classification (MIE is better or not) on \textit{Usefulness} is $-0.03$, which is much lower than those for \textit{Informativeness} (0.1) and \textit{Satisfaction} (0.11). 
One possible reason is that \textit{Usefulness} -- whether the explanations can attract the user to purchase the item -- is a subjective question which varies significantly based on user's preferences and worker's opinions.
In contrast, the questions of whether the explanations provide more information (i.e., \textit{Informativeness}) or increase user's satisfaction on the product search service (i.e., \textit{Satisfaction}) are objective in spite of whether the user purchases the item or not.

To analyze the relation between \textit{Informativeness}, \textit{Usefulness} and \textit{Satisfaction}, we compute the Pearson Coefficient for each pair of labels. 
The coefficients are 0.483 for (\textit{Informativeness}, \textit{Usefulness}), 0.457 for (\textit{Usefulness}, \textit{Satisfaction}), and 0.494 for (\textit{Informativeness}, \textit{Satisfaction}).
Interestingly, we observe that the coefficient between \textit{Usefulness} and \textit{Satisfaction} is the lowest among all pairs.
This may indicate that user's satisfaction on search engines and search explanations is not directly related to whether the explanations encourage the purchases of the items.
Even when a user decide not to purchase an item after seeing the explanations, they may still feel satisfied if the explanations have helped them make more informed decisions. 
The high coefficient between \textit{Informativeness} and \textit{Satisfaction} can also serve as a side evidence for this phenomenon.




\begin{table}[t]
	\centering
	\caption{Crowdsourcing results for explanation evaluation.}
	\scalebox{0.8}{
		\begin{tabular}{ c | c | c | c } 
			\hline
			& \textit{Informativeness \scriptsize{($\kappa=0.10$)}} & \textit{Usefulness} \scriptsize{($\kappa=-0.03$)}& \textit{Satisfaction} \scriptsize{($\kappa=0.11$)} \\ \hline \hline
			MIE wins & 55\% & 43\% & 51\% \\ \hline
			MAE wins & 37\% & 47\% & 35\% \\ \hline \hline
			Equal & 8\% & 11\% & 14\% \\ \hline \hline
		\end{tabular}
	}
	\label{tab:crowd_results}
\end{table}

\begin{figure*}[t]
	\centering
	\includegraphics[width=5.5in]{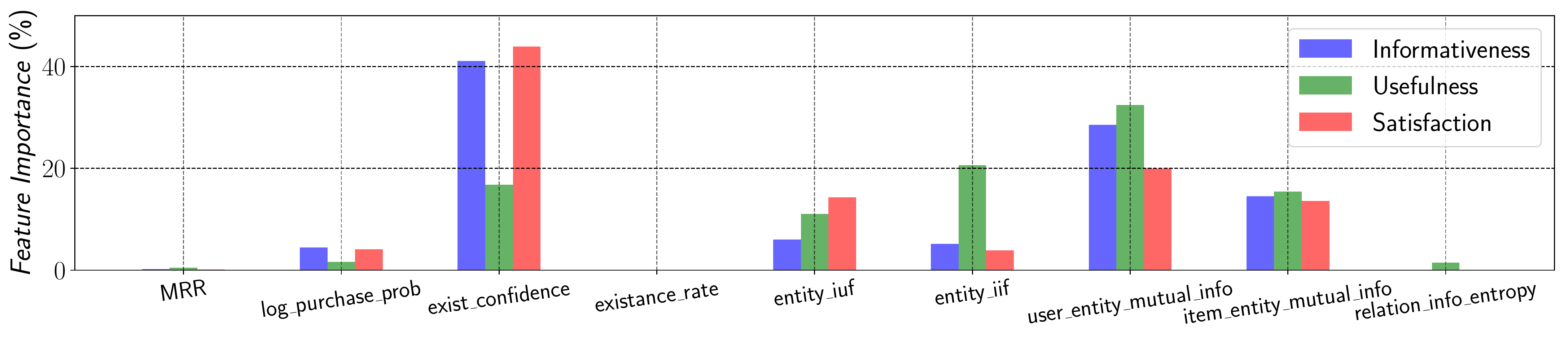}
	\caption{Feature Importance in the GBDT model for explanation performance prediction.}
	\vspace{-10pt}
	\label{fig:feature}
\end{figure*}

\vspace{-5pt}
\subsection{Performance Prediction and Analysis} \label{sec:classification}

To answer \textbf{RQ2}, we propose a performance prediction task for explainable product search by training a classification model to infer user's preferences over search explanations.
By creating such performance prediction models, we want to explore the possibility of evaluating product search explanations without involving human in the loop and analyze the importance of different explanation properties with respect to the effectiveness of search explanations.


\begin{table}[t]
	\centering
	\vspace{-12pt}
	\caption{Descriptions of search explanation features.}
	\scalebox{0.8}{
	\begin{tabular}{ p{8.5cm} } 
	\toprule
	\textbf{Performance}\\
	\textit{MRR}: the MRR of the retrieval model. \\
	\textit{log\_purchase\_prob}: $P(i|u,q)$ in DREM or DREM-HGN.  \\
	\midrule
	\textbf{Fidelity}\\
	\textit{exist\_confidence}: model confidence on the existence of the relations/entities used in explanations ($M(e | u, i)$ for MAE or 1 for MIE).   \\
	\textit{existance\_rate}: percentage of relations/entities used in explanations that are actually observed in the dataset. \\
	\midrule
	\textbf{Novelty}\\
	\textit{entity\_iuf}: inverse user frequency of entities used in explanations.  \\
	\textit{entity\_iif}: inverse item frequency of entities used in explanations.  \\
	\textit{user\_entity\_mutual\_info}: mutual information between users and entities used in explanations in observed data.  \\
	\textit{item\_entity\_mutual\_info}: mutual information between items and entities used in explanations in observed data.  \\
	\textit{relation\_info\_entropy}: the entropy of the distribution of users who have the relations used in explanations.  \\
	\bottomrule
	\end{tabular}
}
	\label{tab:classification_features}
\end{table}

\subsubsection{Feature Design}

The goal of the classification model is to predict user's preferences over an arbitrary pair of result explanations in product search.
To this end, we extract three groups of features to represent each search explanation in the feature space.
They are (1) \textit{Performance} features, which indicate the retrieval effectiveness of the explainable search model that creates the explanation; (2) \textit{Fidelity} features, which indicate whether the information provided by the explanation is correct or trustworthy; and (3) \textit{Novelty} features, which indicate whether the information shown in the explanation is novel or surprising to the user, query, or item. 
For each feature, we compute the maximum, minimum, and mean scores of the entities in each explanation and concatenate the features of all explanations in each group to form the feature vector of the group.
In total, we have 59 features for each explanation group.
Detailed information about each feature can be found in Table~\ref{tab:classification_features}.

To form the feature vector of a pair of explanation groups and avoid introducing biases to the experiment, we concatenated the features of MIE and MAE in both forward and backward orders and create two data points in the classification task for each user-query-item triple. 
Therefore, we have 202 pair of input data and corresponding labels in the performance prediction task.

\subsubsection{Experiment Setup}

We build the performance prediction model with GBDT~\cite{ke2017lightgbm} in LightGBM\footnote{\url{https://github.com/microsoft/LightGBM}}.
We conducted a 5-fold cross validation to predict the pairwise preference of \textit{Informativeness}, \textit{Usefulness} and \textit{Satisfaction}.
For each GBDT, we tuned the maximum tree depth from 5 to 20, leaves number from 10 to 30, minimum leaf data from 10 to 50, and learning rate from 0.1 to 0.5.
The final results are aggregated over all the test folds in cross validation.


\subsubsection{Prediction Results}

\begin{table}[t]
	\centering
	\caption{Explanation performance prediction results.}
	\scalebox{0.8}{
	\begin{tabular}{ r|  c | c | c | c } 
		\hline
		& Total & Correct & Type-1-error & Type-2-error \\ \hline \hline
		\textit{Informativeness} & 202 & 125 & 16 & 61 \\ \hline
		\textit{Usefulness} & 202 & 127 & 24 & 51 \\ \hline
		\textit{Satisfaction} & 202 & 127 & 28 & 47 \\ \hline \hline
	\end{tabular}
}
	\label{tab:classification_results}
\end{table}

Table~\ref{tab:classification_results} depicts the results of our explanation performance prediction experiment.
\textit{Correct} represents the pairs of explanations where the model has correctly predicted their pairwise preferences; \textit{Type-1-error} refers to the pairs where MIE and MAE are equally good while the model predicted that one is better than the other; and \textit{Type-2-error} are cases where MIE or MAE is better than the other while the model predicted otherwise.

\textit{Can we predict the performance of product search explanations without human annotations?}
The overall preference prediction accuracy is 61.9\% for \textit{Informativeness} and 62.9\% for \textit{Usefulness} and \textit{Satisfaction}.
If we allow Type-1-errors, the accuracy would further increase to 69.8\%, 74.8\%, and 76.7\%, respectively.
While such results are far from perfect, they are much better than those produced by a random model and show that it is possible infer user's preferences on product search explanations from their feature representations.
This could serve as an evidence to the potential of automatic product search explanation evaluation in future studies.


\textit{What properties are important for effective product search explanations?}
Intuitively, MIE tends to have higher fidelity because it is directly inferred from the model's internal structure, and MAE may have higher novelty as DREM could extract unobserved entity relations based on soft matching~\cite{ai2019explain}.
As MIE is better than MAE on \textit{Informativeness} and \textit{Satisfaction} while MAE is slightly better than MIE on \textit{Usefulness}, one may expect that fidelity would be more important for the \textit{Informativeness} and \textit{Satisfaction} of product search explanation while novelty may be more important for \textit{Usefulness}.
To examine this hypothesis, we plot the aggregated feature importance in the GBDTs from cross validation based on the total gains of splits which use the feature.
As shown in Figure~\ref{fig:feature}, we observe that fidelity features such as \textit{exist\_confidence} are the most important features for \textit{Informativeness} and \textit{Satisfaction}, which indicates that the reliability of result explanations are important for user's overall satisfaction with the explainable product search engine. 
In contrast, novelty features such as the inverse user/item frequency and the mutual information between user/item and knowledge entities used in explanations are more important for the prediction of \textit{Usefulness}. 
From this perspective, it seems that users are more likely to purchase an item if the search engine could provide some interesting and novel explanations to why it retrieves the corresponding item.

Also, in our experiments, we observe that performance features (i.e., MRR and log\_purchase\_prob) have shown no or minor effect on user's preference over MIE or MAE.
This may because we have filtered out test cases where DREM and DREM-HGN have significant retrieval performance differences (e.g., we only sampled cases where both DREM and DREM-HGN have MRR $\geq$ 0.1 for crowdsourcing), but it may also indicate that users are not sensitive to the retrieval performance of a product search engine when judging the quality of search explanations.
In other words, building explainable product search models that provide effective explanations requires us to rethink of the model design from different perspectives but not simply focusing on the optimization of retrieval performance.





\vspace{-5pt}
\section{Conclusion and Future Work}\label{sec:conclusion}

In this paper, we present the first study that compares model-intrinsic and model-agnostic explanations for explainable product search.
Specifically, we propose a hierarchical gated network as an extension to the state-of-the-art explainable product search model (i.e., DREM), and then conduct a series of experiments to compare and analyze the effectiveness of the post-hoc and pre-hoc search explanations generated by the vanilla DREM and DREM with HGN.

We acknowledge that there are still many limitations of this study such as the template-based explanation generation, the systematic bias introduced by UI and the differences between AMT workers and real product search users.
In future, we will seek for opportunities of online experiments with real product search engines to further analyze the effectiveness of product search explanations and validate the observations in this paper.

\iftrue
\vspace{-5pt}
\section{Acknowledgments}
This work was supported in part by the School of Computing, University of Utah and in part by NSF IIS-2007398. Any opinions, findings and conclusions or recommendations expressed in this material are those of the authors and do not necessarily reflect those of the sponsor.
\fi

\bibliographystyle{ACM-Reference-Format}
\bibliography{sigproc} 


\begin{thebibliography}{76}


\ifx \showCODEN    \undefined \def \showCODEN     #1{\unskip}     \fi
\ifx \showDOI      \undefined \def \showDOI       #1{#1}\fi
\ifx \showISBNx    \undefined \def \showISBNx     #1{\unskip}     \fi
\ifx \showISBNxiii \undefined \def \showISBNxiii  #1{\unskip}     \fi
\ifx \showISSN     \undefined \def \showISSN      #1{\unskip}     \fi
\ifx \showLCCN     \undefined \def \showLCCN      #1{\unskip}     \fi
\ifx \shownote     \undefined \def \shownote      #1{#1}          \fi
\ifx \showarticletitle \undefined \def \showarticletitle #1{#1}   \fi
\ifx \showURL      \undefined \def \showURL       {\relax}        \fi
\providecommand\bibfield[2]{#2}
\providecommand\bibinfo[2]{#2}
\providecommand\natexlab[1]{#1}
\providecommand\showeprint[2][]{arXiv:#2}

\bibitem[\protect\citeauthoryear{Ai, Hill, Vishwanathan, and Croft}{Ai
  et~al\mbox{.}}{2019a}]%
        {ai2019zero}
\bibfield{author}{\bibinfo{person}{Qingyao Ai}, \bibinfo{person}{Daniel Hill},
  \bibinfo{person}{Vishy Vishwanathan}, {and} \bibinfo{person}{W~Bruce Croft}.}
  \bibinfo{year}{2019}\natexlab{a}.
\newblock \showarticletitle{A Zero Attention Model for Personalized Product
  Search}. In \bibinfo{booktitle}{\emph{Proceedings of the 28th ACM
  international on conference on information and knowledge management}}. ACM.
\newblock


\bibitem[\protect\citeauthoryear{Ai, Zhang, Bi, Chen, and Croft}{Ai
  et~al\mbox{.}}{2017}]%
        {ai2017learning}
\bibfield{author}{\bibinfo{person}{Qingyao Ai}, \bibinfo{person}{Yongfeng
  Zhang}, \bibinfo{person}{Keping Bi}, \bibinfo{person}{Xu Chen}, {and}
  \bibinfo{person}{W~Bruce Croft}.} \bibinfo{year}{2017}\natexlab{}.
\newblock \showarticletitle{Learning a hierarchical embedding model for
  personalized product search}. In \bibinfo{booktitle}{\emph{Proceedings of the
  40th International ACM SIGIR Conference on Research and Development in
  Information Retrieval}}. ACM, \bibinfo{pages}{645--654}.
\newblock


\bibitem[\protect\citeauthoryear{Ai, Zhang, Bi, and Croft}{Ai
  et~al\mbox{.}}{2019b}]%
        {ai2019explain}
\bibfield{author}{\bibinfo{person}{Qingyao Ai}, \bibinfo{person}{Yongfeng
  Zhang}, \bibinfo{person}{Keping Bi}, {and} \bibinfo{person}{W~Bruce Croft}.}
  \bibinfo{year}{2019}\natexlab{b}.
\newblock \showarticletitle{Explainable Product Search with a Dynamic Relation
  Embedding Model}.
\newblock \bibinfo{journal}{\emph{ACM Transactions on Information Systems
  (TOIS)}} (\bibinfo{year}{2019}).
\newblock


\bibitem[\protect\citeauthoryear{Andrews, Diederich, and Tickle}{Andrews
  et~al\mbox{.}}{1995}]%
        {andrews1995survey}
\bibfield{author}{\bibinfo{person}{Robert Andrews}, \bibinfo{person}{Joachim
  Diederich}, {and} \bibinfo{person}{Alan~B Tickle}.}
  \bibinfo{year}{1995}\natexlab{}.
\newblock \showarticletitle{Survey and critique of techniques for extracting
  rules from trained artificial neural networks}.
\newblock \bibinfo{journal}{\emph{Knowledge-based systems}}
  \bibinfo{volume}{8}, \bibinfo{number}{6} (\bibinfo{year}{1995}),
  \bibinfo{pages}{373--389}.
\newblock


\bibitem[\protect\citeauthoryear{Aryafar, Guillory, and Hong}{Aryafar
  et~al\mbox{.}}{2017}]%
        {aryafar2017ensemble}
\bibfield{author}{\bibinfo{person}{Kamelia Aryafar}, \bibinfo{person}{Devin
  Guillory}, {and} \bibinfo{person}{Liangjie Hong}.}
  \bibinfo{year}{2017}\natexlab{}.
\newblock \showarticletitle{An ensemble-based approach to click-through rate
  prediction for promoted listings at Etsy}.
\newblock In \bibinfo{booktitle}{\emph{Proceedings of the ADKDD'17}}.
  \bibinfo{pages}{1--6}.
\newblock


\bibitem[\protect\citeauthoryear{Bau, Zhou, Khosla, Oliva, and Torralba}{Bau
  et~al\mbox{.}}{2017}]%
        {bau2017network}
\bibfield{author}{\bibinfo{person}{David Bau}, \bibinfo{person}{Bolei Zhou},
  \bibinfo{person}{Aditya Khosla}, \bibinfo{person}{Aude Oliva}, {and}
  \bibinfo{person}{Antonio Torralba}.} \bibinfo{year}{2017}\natexlab{}.
\newblock \showarticletitle{Network dissection: Quantifying interpretability of
  deep visual representations}. In \bibinfo{booktitle}{\emph{Proceedings of the
  IEEE Conference on Computer Vision and Pattern Recognition}}.
  \bibinfo{pages}{6541--6549}.
\newblock


\bibitem[\protect\citeauthoryear{Bi, Ai, Zhang, and Croft}{Bi
  et~al\mbox{.}}{2019a}]%
        {bi2019negative}
\bibfield{author}{\bibinfo{person}{Keping Bi}, \bibinfo{person}{Qingyao Ai},
  \bibinfo{person}{Yongfeng Zhang}, {and} \bibinfo{person}{W~Bruce Croft}.}
  \bibinfo{year}{2019}\natexlab{a}.
\newblock \showarticletitle{Conversational Product Search Based on Negative
  Feedback}. In \bibinfo{booktitle}{\emph{Proceedings of the 28th ACM
  international on conference on information and knowledge management}}. ACM.
\newblock


\bibitem[\protect\citeauthoryear{Bi, Teo, Dattatreya, Mohan, and Croft}{Bi
  et~al\mbox{.}}{2019b}]%
        {bi2019study}
\bibfield{author}{\bibinfo{person}{Keping Bi}, \bibinfo{person}{Choon~Hui Teo},
  \bibinfo{person}{Yesh Dattatreya}, \bibinfo{person}{Vijai Mohan}, {and}
  \bibinfo{person}{W~Bruce Croft}.} \bibinfo{year}{2019}\natexlab{b}.
\newblock \showarticletitle{A Study of Context Dependencies in Multi-page
  Product Search}. In \bibinfo{booktitle}{\emph{Proceedings of the 28th ACM
  International Conference on Information and Knowledge Management}}.
  \bibinfo{pages}{2333--2336}.
\newblock


\bibitem[\protect\citeauthoryear{Bilgic and Mooney}{Bilgic and Mooney}{2005}]%
        {bilgic2005explaining}
\bibfield{author}{\bibinfo{person}{Mustafa Bilgic} {and}
  \bibinfo{person}{Raymond~J Mooney}.} \bibinfo{year}{2005}\natexlab{}.
\newblock \showarticletitle{Explaining recommendations: Satisfaction vs.
  promotion}. In \bibinfo{booktitle}{\emph{Beyond Personalization Workshop,
  IUI}}, Vol.~\bibinfo{volume}{5}. \bibinfo{pages}{153}.
\newblock


\bibitem[\protect\citeauthoryear{Bordes, Usunier, Garcia-Duran, Weston, and
  Yakhnenko}{Bordes et~al\mbox{.}}{2013}]%
        {bordes2013translating}
\bibfield{author}{\bibinfo{person}{Antoine Bordes}, \bibinfo{person}{Nicolas
  Usunier}, \bibinfo{person}{Alberto Garcia-Duran}, \bibinfo{person}{Jason
  Weston}, {and} \bibinfo{person}{Oksana Yakhnenko}.}
  \bibinfo{year}{2013}\natexlab{}.
\newblock \showarticletitle{Translating embeddings for modeling
  multi-relational data}.
\newblock \bibinfo{journal}{\emph{Advances in neural information processing
  systems}}  \bibinfo{volume}{26} (\bibinfo{year}{2013}),
  \bibinfo{pages}{2787--2795}.
\newblock


\bibitem[\protect\citeauthoryear{Burgess, Higgins, Pal, Matthey, Watters,
  Desjardins, and Lerchner}{Burgess et~al\mbox{.}}{2017}]%
        {burgess2017understanding}
\bibfield{author}{\bibinfo{person}{Christopher~P Burgess},
  \bibinfo{person}{Irina Higgins}, \bibinfo{person}{Arka Pal},
  \bibinfo{person}{Loic Matthey}, \bibinfo{person}{Nick Watters},
  \bibinfo{person}{Guillaume Desjardins}, {and} \bibinfo{person}{Alexander
  Lerchner}.} \bibinfo{year}{2017}\natexlab{}.
\newblock \showarticletitle{Understanding disentangling in $beta$-VAE}.
\newblock \bibinfo{journal}{\emph{Proceedings of the 2017 NIPS Workshop on
  Learning Disentangled Representations}} (\bibinfo{year}{2017}).
\newblock


\bibitem[\protect\citeauthoryear{Burke}{Burke}{2002}]%
        {burke2002hybrid}
\bibfield{author}{\bibinfo{person}{Robin Burke}.}
  \bibinfo{year}{2002}\natexlab{}.
\newblock \showarticletitle{Hybrid recommender systems: Survey and
  experiments}.
\newblock \bibinfo{journal}{\emph{User modeling and user-adapted interaction}}
  \bibinfo{volume}{12}, \bibinfo{number}{4} (\bibinfo{year}{2002}),
  \bibinfo{pages}{331--370}.
\newblock


\bibitem[\protect\citeauthoryear{Carmel, Haramaty, Lazerson, and
  Lewin-Eytan}{Carmel et~al\mbox{.}}{2020a}]%
        {carmel2020multi}
\bibfield{author}{\bibinfo{person}{David Carmel}, \bibinfo{person}{Elad
  Haramaty}, \bibinfo{person}{Arnon Lazerson}, {and} \bibinfo{person}{Liane
  Lewin-Eytan}.} \bibinfo{year}{2020}\natexlab{a}.
\newblock \showarticletitle{Multi-Objective Ranking Optimization for Product
  Search Using Stochastic Label Aggregation}. In
  \bibinfo{booktitle}{\emph{Proceedings of The Web Conference 2020}}.
  \bibinfo{pages}{373--383}.
\newblock


\bibitem[\protect\citeauthoryear{Carmel, Haramaty, Lazerson, Lewin-Eytan, and
  Maarek}{Carmel et~al\mbox{.}}{2020b}]%
        {10.1145/3336191.3371780}
\bibfield{author}{\bibinfo{person}{David Carmel}, \bibinfo{person}{Elad
  Haramaty}, \bibinfo{person}{Arnon Lazerson}, \bibinfo{person}{Liane
  Lewin-Eytan}, {and} \bibinfo{person}{Yoelle Maarek}.}
  \bibinfo{year}{2020}\natexlab{b}.
\newblock \showarticletitle{Why Do People Buy Seemingly Irrelevant Items in
  Voice Product Search? On the Relation between Product Relevance and Customer
  Satisfaction in ECommerce}. In \bibinfo{booktitle}{\emph{Proceedings of the
  13th International Conference on Web Search and Data Mining}} (Houston, TX,
  USA) \emph{(\bibinfo{series}{WSDM '20})}. \bibinfo{publisher}{Association for
  Computing Machinery}, \bibinfo{address}{New York, NY, USA},
  \bibinfo{pages}{79–87}.
\newblock
\showISBNx{9781450368223}
\urldef\tempurl%
\url{https://doi.org/10.1145/3336191.3371780}
\showDOI{\tempurl}


\bibitem[\protect\citeauthoryear{Cramer, Evers, Ramlal, Van~Someren, Rutledge,
  Stash, Aroyo, and Wielinga}{Cramer et~al\mbox{.}}{2008}]%
        {cramer2008effects}
\bibfield{author}{\bibinfo{person}{Henriette Cramer}, \bibinfo{person}{Vanessa
  Evers}, \bibinfo{person}{Satyan Ramlal}, \bibinfo{person}{Maarten
  Van~Someren}, \bibinfo{person}{Lloyd Rutledge}, \bibinfo{person}{Natalia
  Stash}, \bibinfo{person}{Lora Aroyo}, {and} \bibinfo{person}{Bob Wielinga}.}
  \bibinfo{year}{2008}\natexlab{}.
\newblock \showarticletitle{The effects of transparency on trust in and
  acceptance of a content-based art recommender}.
\newblock \bibinfo{journal}{\emph{User Modeling and User-Adapted Interaction}}
  \bibinfo{volume}{18}, \bibinfo{number}{5} (\bibinfo{year}{2008}),
  \bibinfo{pages}{455}.
\newblock


\bibitem[\protect\citeauthoryear{Du, Liu, and Hu}{Du et~al\mbox{.}}{2018}]%
        {du2018techniques}
\bibfield{author}{\bibinfo{person}{Mengnan Du}, \bibinfo{person}{Ninghao Liu},
  {and} \bibinfo{person}{Xia Hu}.} \bibinfo{year}{2018}\natexlab{}.
\newblock \showarticletitle{Techniques for interpretable machine learning}.
\newblock \bibinfo{journal}{\emph{Commun. ACM}} (\bibinfo{year}{2018}).
\newblock


\bibitem[\protect\citeauthoryear{Duan, Zhai, Cheng, and Gattani}{Duan
  et~al\mbox{.}}{2013a}]%
        {duan2013probabilistic}
\bibfield{author}{\bibinfo{person}{Huizhong Duan}, \bibinfo{person}{ChengXiang
  Zhai}, \bibinfo{person}{Jinxing Cheng}, {and} \bibinfo{person}{Abhishek
  Gattani}.} \bibinfo{year}{2013}\natexlab{a}.
\newblock \showarticletitle{A probabilistic mixture model for mining and
  analyzing product search log}. In \bibinfo{booktitle}{\emph{Proceedings of
  the 22nd ACM international conference on Information \& Knowledge
  Management}}. \bibinfo{pages}{2179--2188}.
\newblock


\bibitem[\protect\citeauthoryear{Duan, Zhai, Cheng, and Gattani}{Duan
  et~al\mbox{.}}{2013b}]%
        {duan2013supporting}
\bibfield{author}{\bibinfo{person}{Huizhong Duan}, \bibinfo{person}{ChengXiang
  Zhai}, \bibinfo{person}{Jinxing Cheng}, {and} \bibinfo{person}{Abhishek
  Gattani}.} \bibinfo{year}{2013}\natexlab{b}.
\newblock \showarticletitle{Supporting keyword search in product database: a
  probabilistic approach}.
\newblock \bibinfo{journal}{\emph{Proceedings of the VLDB Endowment}}
  \bibinfo{volume}{6}, \bibinfo{number}{14} (\bibinfo{year}{2013}),
  \bibinfo{pages}{1786--1797}.
\newblock


\bibitem[\protect\citeauthoryear{Fernando, Singh, and Anand}{Fernando
  et~al\mbox{.}}{2019}]%
        {fernando2019study}
\bibfield{author}{\bibinfo{person}{Zeon~Trevor Fernando},
  \bibinfo{person}{Jaspreet Singh}, {and} \bibinfo{person}{Avishek Anand}.}
  \bibinfo{year}{2019}\natexlab{}.
\newblock \showarticletitle{A study on the Interpretability of Neural Retrieval
  Models using DeepSHAP}. In \bibinfo{booktitle}{\emph{Proceedings of the 42nd
  International ACM SIGIR Conference on Research and Development in Information
  Retrieval}}. \bibinfo{pages}{1005--1008}.
\newblock


\bibitem[\protect\citeauthoryear{Frankle and Carbin}{Frankle and
  Carbin}{2019}]%
        {frankle2018lottery}
\bibfield{author}{\bibinfo{person}{Jonathan Frankle} {and}
  \bibinfo{person}{Michael Carbin}.} \bibinfo{year}{2019}\natexlab{}.
\newblock \showarticletitle{The lottery ticket hypothesis: Finding sparse,
  trainable neural networks}.
\newblock \bibinfo{journal}{\emph{ICLR}} (\bibinfo{year}{2019}).
\newblock


\bibitem[\protect\citeauthoryear{Fu}{Fu}{1994}]%
        {fu1994rule}
\bibfield{author}{\bibinfo{person}{LiMin Fu}.} \bibinfo{year}{1994}\natexlab{}.
\newblock \showarticletitle{Rule generation from neural networks}.
\newblock \bibinfo{journal}{\emph{IEEE Transactions on Systems, Man, and
  Cybernetics}} \bibinfo{volume}{24}, \bibinfo{number}{8}
  (\bibinfo{year}{1994}), \bibinfo{pages}{1114--1124}.
\newblock


\bibitem[\protect\citeauthoryear{Gilpin, Bau, Yuan, Bajwa, Specter, and
  Kagal}{Gilpin et~al\mbox{.}}{2018}]%
        {gilpin2018explaining}
\bibfield{author}{\bibinfo{person}{Leilani~H Gilpin}, \bibinfo{person}{David
  Bau}, \bibinfo{person}{Ben~Z Yuan}, \bibinfo{person}{Ayesha Bajwa},
  \bibinfo{person}{Michael Specter}, {and} \bibinfo{person}{Lalana Kagal}.}
  \bibinfo{year}{2018}\natexlab{}.
\newblock \showarticletitle{Explaining explanations: An overview of
  interpretability of machine learning}. In \bibinfo{booktitle}{\emph{2018 IEEE
  5th International Conference on data science and advanced analytics (DSAA)}}.
  IEEE, \bibinfo{pages}{80--89}.
\newblock


\bibitem[\protect\citeauthoryear{Guo, Fan, Pang, Yang, Ai, Zamani, Wu, Croft,
  and Cheng}{Guo et~al\mbox{.}}{2019b}]%
        {guo2019deep}
\bibfield{author}{\bibinfo{person}{Jiafeng Guo}, \bibinfo{person}{Yixing Fan},
  \bibinfo{person}{Liang Pang}, \bibinfo{person}{Liu Yang},
  \bibinfo{person}{Qingyao Ai}, \bibinfo{person}{Hamed Zamani},
  \bibinfo{person}{Chen Wu}, \bibinfo{person}{W~Bruce Croft}, {and}
  \bibinfo{person}{Xueqi Cheng}.} \bibinfo{year}{2019}\natexlab{b}.
\newblock \showarticletitle{A deep look into neural ranking models for
  information retrieval}.
\newblock \bibinfo{journal}{\emph{arXiv preprint arXiv:1903.06902}}
  (\bibinfo{year}{2019}).
\newblock


\bibitem[\protect\citeauthoryear{Guo, Cheng, Nie, Wang, Ma, and
  Kankanhalli}{Guo et~al\mbox{.}}{2019a}]%
        {guo2019attentive}
\bibfield{author}{\bibinfo{person}{Yangyang Guo}, \bibinfo{person}{Zhiyong
  Cheng}, \bibinfo{person}{Liqiang Nie}, \bibinfo{person}{Yinglong Wang},
  \bibinfo{person}{Jun Ma}, {and} \bibinfo{person}{Mohan Kankanhalli}.}
  \bibinfo{year}{2019}\natexlab{a}.
\newblock \showarticletitle{Attentive long short-term preference modeling for
  personalized product search}.
\newblock \bibinfo{journal}{\emph{ACM Transactions on Information Systems
  (TOIS)}} \bibinfo{volume}{37}, \bibinfo{number}{2} (\bibinfo{year}{2019}),
  \bibinfo{pages}{1--27}.
\newblock


\bibitem[\protect\citeauthoryear{Guo, Cheng, Nie, Xu, and Kankanhalli}{Guo
  et~al\mbox{.}}{2018}]%
        {guo2018multi}
\bibfield{author}{\bibinfo{person}{Yangyang Guo}, \bibinfo{person}{Zhiyong
  Cheng}, \bibinfo{person}{Liqiang Nie}, \bibinfo{person}{Xin-Shun Xu}, {and}
  \bibinfo{person}{Mohan Kankanhalli}.} \bibinfo{year}{2018}\natexlab{}.
\newblock \showarticletitle{Multi-modal preference modeling for product
  search}. In \bibinfo{booktitle}{\emph{Proceedings of the 26th ACM
  international conference on Multimedia}}. \bibinfo{pages}{1865--1873}.
\newblock


\bibitem[\protect\citeauthoryear{Herlocker}{Herlocker}{2000}]%
        {herlocker2000understanding}
\bibfield{author}{\bibinfo{person}{Jonathan~Lee Herlocker}.}
  \bibinfo{year}{2000}\natexlab{}.
\newblock \bibinfo{booktitle}{\emph{Understanding and improving automated
  collaborative filtering systems}}.
\newblock \bibinfo{publisher}{Citeseer}.
\newblock


\bibitem[\protect\citeauthoryear{Herlocker, Konstan, and Riedl}{Herlocker
  et~al\mbox{.}}{2000}]%
        {herlocker2000explaining}
\bibfield{author}{\bibinfo{person}{Jonathan~L Herlocker},
  \bibinfo{person}{Joseph~A Konstan}, {and} \bibinfo{person}{John Riedl}.}
  \bibinfo{year}{2000}\natexlab{}.
\newblock \showarticletitle{Explaining collaborative filtering
  recommendations}. In \bibinfo{booktitle}{\emph{Proceedings of the 2000 ACM
  conference on Computer supported cooperative work}}. ACM,
  \bibinfo{pages}{241--250}.
\newblock


\bibitem[\protect\citeauthoryear{Higgins, Matthey, Pal, Burgess, Glorot,
  Botvinick, Mohamed, and Lerchner}{Higgins et~al\mbox{.}}{2017}]%
        {higgins2017beta}
\bibfield{author}{\bibinfo{person}{Irina Higgins}, \bibinfo{person}{Loic
  Matthey}, \bibinfo{person}{Arka Pal}, \bibinfo{person}{Christopher Burgess},
  \bibinfo{person}{Xavier Glorot}, \bibinfo{person}{Matthew Botvinick},
  \bibinfo{person}{Shakir Mohamed}, {and} \bibinfo{person}{Alexander
  Lerchner}.} \bibinfo{year}{2017}\natexlab{}.
\newblock \showarticletitle{beta-VAE: Learning Basic Visual Concepts with a
  Constrained Variational Framework.}
\newblock \bibinfo{journal}{\emph{ICLR}} \bibinfo{volume}{2},
  \bibinfo{number}{5} (\bibinfo{year}{2017}), \bibinfo{pages}{6}.
\newblock


\bibitem[\protect\citeauthoryear{Hu, Da, Zeng, Yu, and Xu}{Hu
  et~al\mbox{.}}{2018}]%
        {hu2018reinforcement}
\bibfield{author}{\bibinfo{person}{Yujing Hu}, \bibinfo{person}{Qing Da},
  \bibinfo{person}{Anxiang Zeng}, \bibinfo{person}{Yang Yu}, {and}
  \bibinfo{person}{Yinghui Xu}.} \bibinfo{year}{2018}\natexlab{}.
\newblock \showarticletitle{Reinforcement learning to rank in e-commerce search
  engine: Formalization, analysis, and application}. In
  \bibinfo{booktitle}{\emph{Proceedings of the 24th ACM SIGKDD International
  Conference on Knowledge Discovery \& Data Mining}}.
  \bibinfo{pages}{368--377}.
\newblock


\bibitem[\protect\citeauthoryear{Jain and Wallace}{Jain and Wallace}{2019}]%
        {jain2019attention}
\bibfield{author}{\bibinfo{person}{Sarthak Jain} {and} \bibinfo{person}{Byron~C
  Wallace}.} \bibinfo{year}{2019}\natexlab{}.
\newblock \showarticletitle{Attention is not explanation}.
\newblock \bibinfo{journal}{\emph{Proceedings of the 2019 Conference of the
  North American Chapter of the Association for Computational Linguistics}}
  (\bibinfo{year}{2019}).
\newblock


\bibitem[\protect\citeauthoryear{Joachims, Granka, Pan, Hembrooke, and
  Gay}{Joachims et~al\mbox{.}}{2017}]%
        {joachims2017accurately}
\bibfield{author}{\bibinfo{person}{Thorsten Joachims}, \bibinfo{person}{Laura
  Granka}, \bibinfo{person}{Bing Pan}, \bibinfo{person}{Helene Hembrooke},
  {and} \bibinfo{person}{Geri Gay}.} \bibinfo{year}{2017}\natexlab{}.
\newblock \showarticletitle{Accurately interpreting clickthrough data as
  implicit feedback}. In \bibinfo{booktitle}{\emph{ACM SIGIR Forum}},
  Vol.~\bibinfo{volume}{51}. Acm New York, NY, USA, \bibinfo{pages}{4--11}.
\newblock


\bibitem[\protect\citeauthoryear{Karmaker~Santu, Sondhi, and
  Zhai}{Karmaker~Santu et~al\mbox{.}}{2017}]%
        {karmaker2017application}
\bibfield{author}{\bibinfo{person}{Shubhra~Kanti Karmaker~Santu},
  \bibinfo{person}{Parikshit Sondhi}, {and} \bibinfo{person}{ChengXiang Zhai}.}
  \bibinfo{year}{2017}\natexlab{}.
\newblock \showarticletitle{On application of learning to rank for e-commerce
  search}. In \bibinfo{booktitle}{\emph{Proceedings of the 40th International
  ACM SIGIR Conference on Research and Development in Information Retrieval}}.
  \bibinfo{pages}{475--484}.
\newblock


\bibitem[\protect\citeauthoryear{Ke, Meng, Finley, Wang, Chen, Ma, Ye, and
  Liu}{Ke et~al\mbox{.}}{2017}]%
        {ke2017lightgbm}
\bibfield{author}{\bibinfo{person}{Guolin Ke}, \bibinfo{person}{Qi Meng},
  \bibinfo{person}{Thomas Finley}, \bibinfo{person}{Taifeng Wang},
  \bibinfo{person}{Wei Chen}, \bibinfo{person}{Weidong Ma},
  \bibinfo{person}{Qiwei Ye}, {and} \bibinfo{person}{Tie-Yan Liu}.}
  \bibinfo{year}{2017}\natexlab{}.
\newblock \showarticletitle{Lightgbm: A highly efficient gradient boosting
  decision tree}. In \bibinfo{booktitle}{\emph{Advances in neural information
  processing systems}}. \bibinfo{pages}{3146--3154}.
\newblock


\bibitem[\protect\citeauthoryear{Levy and Goldberg}{Levy and Goldberg}{2014}]%
        {levy2014neural}
\bibfield{author}{\bibinfo{person}{Omer Levy} {and} \bibinfo{person}{Yoav
  Goldberg}.} \bibinfo{year}{2014}\natexlab{}.
\newblock \showarticletitle{Neural word embedding as implicit matrix
  factorization}.
\newblock \bibinfo{journal}{\emph{Advances in neural information processing
  systems}}  \bibinfo{volume}{27} (\bibinfo{year}{2014}),
  \bibinfo{pages}{2177--2185}.
\newblock


\bibitem[\protect\citeauthoryear{Lim, Wang, and Wang}{Lim
  et~al\mbox{.}}{2013}]%
        {lim2013semantic}
\bibfield{author}{\bibinfo{person}{Lipyeow Lim}, \bibinfo{person}{Haixun Wang},
  {and} \bibinfo{person}{Min Wang}.} \bibinfo{year}{2013}\natexlab{}.
\newblock \showarticletitle{Semantic queries by example}. In
  \bibinfo{booktitle}{\emph{Proceedings of the 16th International Conference on
  Extending Database Technology}}. ACM, \bibinfo{pages}{347--358}.
\newblock


\bibitem[\protect\citeauthoryear{Lipton}{Lipton}{2018}]%
        {lipton2018mythos}
\bibfield{author}{\bibinfo{person}{Zachary~C Lipton}.}
  \bibinfo{year}{2018}\natexlab{}.
\newblock \showarticletitle{The mythos of model interpretability}.
\newblock \bibinfo{journal}{\emph{Commun. ACM}} \bibinfo{volume}{61},
  \bibinfo{number}{10} (\bibinfo{year}{2018}), \bibinfo{pages}{36--43}.
\newblock


\bibitem[\protect\citeauthoryear{Liu, Gu, Cong, and Zhang}{Liu
  et~al\mbox{.}}{2020}]%
        {liu2020structural}
\bibfield{author}{\bibinfo{person}{Shang Liu}, \bibinfo{person}{Wanli Gu},
  \bibinfo{person}{Gao Cong}, {and} \bibinfo{person}{Fuzheng Zhang}.}
  \bibinfo{year}{2020}\natexlab{}.
\newblock \showarticletitle{Structural Relationship Representation Learning
  with Graph Embedding for Personalized Product Search}. In
  \bibinfo{booktitle}{\emph{Proceedings of the 29th ACM International
  Conference on Information \& Knowledge Management}}.
  \bibinfo{pages}{915--924}.
\newblock


\bibitem[\protect\citeauthoryear{Locatello, Bauer, Lucic, Gelly, Sch{\"o}lkopf,
  and Bachem}{Locatello et~al\mbox{.}}{2018}]%
        {locatello2018challenging}
\bibfield{author}{\bibinfo{person}{Francesco Locatello},
  \bibinfo{person}{Stefan Bauer}, \bibinfo{person}{Mario Lucic},
  \bibinfo{person}{Sylvain Gelly}, \bibinfo{person}{Bernhard Sch{\"o}lkopf},
  {and} \bibinfo{person}{Olivier Bachem}.} \bibinfo{year}{2018}\natexlab{}.
\newblock \showarticletitle{Challenging common assumptions in the unsupervised
  learning of disentangled representations}.
\newblock \bibinfo{journal}{\emph{Proceedings of the 36th International
  Conference on Machine Learning}} (\bibinfo{year}{2018}).
\newblock


\bibitem[\protect\citeauthoryear{Lundberg and Lee}{Lundberg and Lee}{2017}]%
        {lundberg2017unified}
\bibfield{author}{\bibinfo{person}{Scott~M Lundberg} {and}
  \bibinfo{person}{Su-In Lee}.} \bibinfo{year}{2017}\natexlab{}.
\newblock \showarticletitle{A Unified Approach to Interpreting Model
  Predictions}.
\newblock \bibinfo{journal}{\emph{Advances in Neural Information Processing
  Systems}}  \bibinfo{volume}{30} (\bibinfo{year}{2017}),
  \bibinfo{pages}{4765--4774}.
\newblock


\bibitem[\protect\citeauthoryear{Luo, Xiong, Liu, and Sun}{Luo
  et~al\mbox{.}}{2019}]%
        {luo2019adaptive}
\bibfield{author}{\bibinfo{person}{Liangchen Luo}, \bibinfo{person}{Yuanhao
  Xiong}, \bibinfo{person}{Yan Liu}, {and} \bibinfo{person}{Xu Sun}.}
  \bibinfo{year}{2019}\natexlab{}.
\newblock \showarticletitle{Adaptive gradient methods with dynamic bound of
  learning rate}.
\newblock \bibinfo{journal}{\emph{arXiv preprint arXiv:1902.09843}}
  (\bibinfo{year}{2019}).
\newblock


\bibitem[\protect\citeauthoryear{McAuley and Leskovec}{McAuley and
  Leskovec}{2013}]%
        {mcauley2013hidden}
\bibfield{author}{\bibinfo{person}{Julian McAuley} {and} \bibinfo{person}{Jure
  Leskovec}.} \bibinfo{year}{2013}\natexlab{}.
\newblock \showarticletitle{Hidden factors and hidden topics: understanding
  rating dimensions with review text}. In \bibinfo{booktitle}{\emph{Proceedings
  of the 7th ACM conference on Recommender systems}}. ACM,
  \bibinfo{pages}{165--172}.
\newblock


\bibitem[\protect\citeauthoryear{McAuley, Targett, Shi, and Van
  Den~Hengel}{McAuley et~al\mbox{.}}{2015}]%
        {mcauley2015image}
\bibfield{author}{\bibinfo{person}{Julian McAuley},
  \bibinfo{person}{Christopher Targett}, \bibinfo{person}{Qinfeng Shi}, {and}
  \bibinfo{person}{Anton Van Den~Hengel}.} \bibinfo{year}{2015}\natexlab{}.
\newblock \showarticletitle{Image-based recommendations on styles and
  substitutes}. In \bibinfo{booktitle}{\emph{Proceedings of the 38th
  International ACM SIGIR Conference on Research and Development in Information
  Retrieval}}. ACM, \bibinfo{pages}{43--52}.
\newblock


\bibitem[\protect\citeauthoryear{Mikolov, Chen, Corrado, and Dean}{Mikolov
  et~al\mbox{.}}{2013}]%
        {mikolov2013efficient}
\bibfield{author}{\bibinfo{person}{Tomas Mikolov}, \bibinfo{person}{Kai Chen},
  \bibinfo{person}{Greg Corrado}, {and} \bibinfo{person}{Jeffrey Dean}.}
  \bibinfo{year}{2013}\natexlab{}.
\newblock \showarticletitle{Efficient estimation of word representations in
  vector space}.
\newblock \bibinfo{journal}{\emph{arXiv preprint arXiv:1301.3781}}
  (\bibinfo{year}{2013}).
\newblock


\bibitem[\protect\citeauthoryear{Mitra, Craswell, et~al\mbox{.}}{Mitra
  et~al\mbox{.}}{2018}]%
        {mitra2018introduction}
\bibfield{author}{\bibinfo{person}{Bhaskar Mitra}, \bibinfo{person}{Nick
  Craswell}, {et~al\mbox{.}}} \bibinfo{year}{2018}\natexlab{}.
\newblock \showarticletitle{An introduction to neural information retrieval}.
\newblock \bibinfo{journal}{\emph{Foundations and Trends{\textregistered} in
  Information Retrieval}} \bibinfo{volume}{13}, \bibinfo{number}{1}
  (\bibinfo{year}{2018}), \bibinfo{pages}{1--126}.
\newblock


\bibitem[\protect\citeauthoryear{Nguyen, Dosovitskiy, Yosinski, Brox, and
  Clune}{Nguyen et~al\mbox{.}}{2016}]%
        {nguyen2016synthesizing}
\bibfield{author}{\bibinfo{person}{Anh Nguyen}, \bibinfo{person}{Alexey
  Dosovitskiy}, \bibinfo{person}{Jason Yosinski}, \bibinfo{person}{Thomas
  Brox}, {and} \bibinfo{person}{Jeff Clune}.} \bibinfo{year}{2016}\natexlab{}.
\newblock \showarticletitle{Synthesizing the preferred inputs for neurons in
  neural networks via deep generator networks}. In
  \bibinfo{booktitle}{\emph{Advances in Neural Information Processing
  Systems}}. \bibinfo{pages}{3387--3395}.
\newblock


\bibitem[\protect\citeauthoryear{Nurmi, Lagerspetz, Buntine, Flor{\'e}en, and
  Kukkonen}{Nurmi et~al\mbox{.}}{2008}]%
        {nurmi2008product}
\bibfield{author}{\bibinfo{person}{Petteri Nurmi}, \bibinfo{person}{Eemil
  Lagerspetz}, \bibinfo{person}{Wray Buntine}, \bibinfo{person}{Patrik
  Flor{\'e}en}, {and} \bibinfo{person}{Joonas Kukkonen}.}
  \bibinfo{year}{2008}\natexlab{}.
\newblock \showarticletitle{Product retrieval for grocery stores}. In
  \bibinfo{booktitle}{\emph{Proceedings of the 31st annual international ACM
  SIGIR conference on Research and development in information retrieval}}.
  \bibinfo{pages}{781--782}.
\newblock


\bibitem[\protect\citeauthoryear{Peake and Wang}{Peake and Wang}{2018}]%
        {peake2018explanation}
\bibfield{author}{\bibinfo{person}{Georgina Peake} {and} \bibinfo{person}{Jun
  Wang}.} \bibinfo{year}{2018}\natexlab{}.
\newblock \showarticletitle{Explanation mining: Post hoc interpretability of
  latent factor models for recommendation systems}. In
  \bibinfo{booktitle}{\emph{Proceedings of the 24th ACM SIGKDD International
  Conference on Knowledge Discovery \& Data Mining}}. ACM,
  \bibinfo{pages}{2060--2069}.
\newblock


\bibitem[\protect\citeauthoryear{Ponte and Croft}{Ponte and Croft}{1998}]%
        {ponte1998language}
\bibfield{author}{\bibinfo{person}{Jay~Michael Ponte} {and}
  \bibinfo{person}{W~Bruce Croft}.} \bibinfo{year}{1998}\natexlab{}.
\newblock \emph{\bibinfo{title}{A language modeling approach to information
  retrieval}}.
\newblock \bibinfo{thesistype}{Ph.D. Dissertation}. \bibinfo{school}{University
  of Massachusetts at Amherst}.
\newblock


\bibitem[\protect\citeauthoryear{Ribeiro, Singh, and Guestrin}{Ribeiro
  et~al\mbox{.}}{2016}]%
        {ribeiro2016should}
\bibfield{author}{\bibinfo{person}{Marco~Tulio Ribeiro},
  \bibinfo{person}{Sameer Singh}, {and} \bibinfo{person}{Carlos Guestrin}.}
  \bibinfo{year}{2016}\natexlab{}.
\newblock \showarticletitle{Why should i trust you?: Explaining the predictions
  of any classifier}. In \bibinfo{booktitle}{\emph{Proceedings of the 22nd ACM
  SIGKDD international conference on knowledge discovery and data mining}}.
  ACM, \bibinfo{pages}{1135--1144}.
\newblock


\bibitem[\protect\citeauthoryear{Robertson, Zaragoza, et~al\mbox{.}}{Robertson
  et~al\mbox{.}}{2009}]%
        {robertson2009probabilistic}
\bibfield{author}{\bibinfo{person}{Stephen Robertson}, \bibinfo{person}{Hugo
  Zaragoza}, {et~al\mbox{.}}} \bibinfo{year}{2009}\natexlab{}.
\newblock \showarticletitle{The probabilistic relevance framework: BM25 and
  beyond}.
\newblock \bibinfo{journal}{\emph{Foundations and Trends{\textregistered} in
  Information Retrieval}} \bibinfo{volume}{3}, \bibinfo{number}{4}
  (\bibinfo{year}{2009}), \bibinfo{pages}{333--389}.
\newblock


\bibitem[\protect\citeauthoryear{Rowley}{Rowley}{2000}]%
        {rowley2000product}
\bibfield{author}{\bibinfo{person}{Jennifer Rowley}.}
  \bibinfo{year}{2000}\natexlab{}.
\newblock \showarticletitle{Product search in e-shopping: a review and research
  propositions}.
\newblock \bibinfo{journal}{\emph{Journal of consumer marketing}}
  (\bibinfo{year}{2000}).
\newblock


\bibitem[\protect\citeauthoryear{Schmitz, Aldrich, and Gouws}{Schmitz
  et~al\mbox{.}}{1999}]%
        {schmitz1999ann}
\bibfield{author}{\bibinfo{person}{Gregor~PJ Schmitz}, \bibinfo{person}{Chris
  Aldrich}, {and} \bibinfo{person}{Francois~S Gouws}.}
  \bibinfo{year}{1999}\natexlab{}.
\newblock \showarticletitle{ANN-DT: an algorithm for extraction of decision
  trees from artificial neural networks}.
\newblock \bibinfo{journal}{\emph{IEEE Transactions on Neural Networks}}
  \bibinfo{volume}{10}, \bibinfo{number}{6} (\bibinfo{year}{1999}),
  \bibinfo{pages}{1392--1401}.
\newblock


\bibitem[\protect\citeauthoryear{Sharif~Razavian, Azizpour, Sullivan, and
  Carlsson}{Sharif~Razavian et~al\mbox{.}}{2014}]%
        {sharif2014cnn}
\bibfield{author}{\bibinfo{person}{Ali Sharif~Razavian},
  \bibinfo{person}{Hossein Azizpour}, \bibinfo{person}{Josephine Sullivan},
  {and} \bibinfo{person}{Stefan Carlsson}.} \bibinfo{year}{2014}\natexlab{}.
\newblock \showarticletitle{CNN features off-the-shelf: an astounding baseline
  for recognition}. In \bibinfo{booktitle}{\emph{Proceedings of the IEEE
  conference on computer vision and pattern recognition workshops}}.
  \bibinfo{pages}{806--813}.
\newblock


\bibitem[\protect\citeauthoryear{Simonyan, Vedaldi, and Zisserman}{Simonyan
  et~al\mbox{.}}{2013}]%
        {simonyan2013deep}
\bibfield{author}{\bibinfo{person}{Karen Simonyan}, \bibinfo{person}{Andrea
  Vedaldi}, {and} \bibinfo{person}{Andrew Zisserman}.}
  \bibinfo{year}{2013}\natexlab{}.
\newblock \showarticletitle{Deep inside convolutional networks: Visualising
  image classification models and saliency maps}.
\newblock \bibinfo{journal}{\emph{arXiv preprint arXiv:1312.6034}}
  (\bibinfo{year}{2013}).
\newblock


\bibitem[\protect\citeauthoryear{Singh and Anand}{Singh and Anand}{2019}]%
        {singh2019exs}
\bibfield{author}{\bibinfo{person}{Jaspreet Singh} {and}
  \bibinfo{person}{Avishek Anand}.} \bibinfo{year}{2019}\natexlab{}.
\newblock \showarticletitle{Exs: Explainable search using local model agnostic
  interpretability}. In \bibinfo{booktitle}{\emph{Proceedings of the Twelfth
  ACM International Conference on Web Search and Data Mining}}.
  \bibinfo{pages}{770--773}.
\newblock


\bibitem[\protect\citeauthoryear{Singh and Anand}{Singh and Anand}{2020}]%
        {singh2020model}
\bibfield{author}{\bibinfo{person}{Jaspreet Singh} {and}
  \bibinfo{person}{Avishek Anand}.} \bibinfo{year}{2020}\natexlab{}.
\newblock \showarticletitle{Model agnostic interpretability of rankers via
  intent modelling}. In \bibinfo{booktitle}{\emph{Proceedings of the 2020
  Conference on Fairness, Accountability, and Transparency}}.
  \bibinfo{pages}{618--628}.
\newblock


\bibitem[\protect\citeauthoryear{Smucker, Allan, and Carterette}{Smucker
  et~al\mbox{.}}{2007}]%
        {smucker2007comparison}
\bibfield{author}{\bibinfo{person}{Mark~D Smucker}, \bibinfo{person}{James
  Allan}, {and} \bibinfo{person}{Ben Carterette}.}
  \bibinfo{year}{2007}\natexlab{}.
\newblock \showarticletitle{A comparison of statistical significance tests for
  information retrieval evaluation}. In \bibinfo{booktitle}{\emph{Proceedings
  of the sixteenth ACM conference on Conference on information and knowledge
  management}}. \bibinfo{pages}{623--632}.
\newblock


\bibitem[\protect\citeauthoryear{Tintarev and Masthoff}{Tintarev and
  Masthoff}{2007a}]%
        {tintarev2007effective}
\bibfield{author}{\bibinfo{person}{Nava Tintarev} {and} \bibinfo{person}{Judith
  Masthoff}.} \bibinfo{year}{2007}\natexlab{a}.
\newblock \showarticletitle{Effective explanations of recommendations:
  user-centered design}. In \bibinfo{booktitle}{\emph{Proceedings of the 2007
  ACM conference on Recommender systems}}. ACM, \bibinfo{pages}{153--156}.
\newblock


\bibitem[\protect\citeauthoryear{Tintarev and Masthoff}{Tintarev and
  Masthoff}{2007b}]%
        {tintarev2007survey}
\bibfield{author}{\bibinfo{person}{Nava Tintarev} {and} \bibinfo{person}{Judith
  Masthoff}.} \bibinfo{year}{2007}\natexlab{b}.
\newblock \showarticletitle{A survey of explanations in recommender systems}.
  In \bibinfo{booktitle}{\emph{2007 IEEE 23rd international conference on data
  engineering workshop}}. IEEE, \bibinfo{pages}{801--810}.
\newblock


\bibitem[\protect\citeauthoryear{Van~den Oord, Dieleman, and Schrauwen}{Van~den
  Oord et~al\mbox{.}}{2013}]%
        {van2013deep}
\bibfield{author}{\bibinfo{person}{Aaron Van~den Oord}, \bibinfo{person}{Sander
  Dieleman}, {and} \bibinfo{person}{Benjamin Schrauwen}.}
  \bibinfo{year}{2013}\natexlab{}.
\newblock \showarticletitle{Deep content-based music recommendation}. In
  \bibinfo{booktitle}{\emph{Advances in neural information processing
  systems}}. \bibinfo{pages}{2643--2651}.
\newblock


\bibitem[\protect\citeauthoryear{Van~Gysel, de~Rijke, and Kanoulas}{Van~Gysel
  et~al\mbox{.}}{2016}]%
        {van2016learning}
\bibfield{author}{\bibinfo{person}{Christophe Van~Gysel},
  \bibinfo{person}{Maarten de Rijke}, {and} \bibinfo{person}{Evangelos
  Kanoulas}.} \bibinfo{year}{2016}\natexlab{}.
\newblock \showarticletitle{Learning latent vector spaces for product search}.
  In \bibinfo{booktitle}{\emph{Proceedings of the 25th ACM International on
  Conference on Information and Knowledge Management}}.
  \bibinfo{pages}{165--174}.
\newblock


\bibitem[\protect\citeauthoryear{Vaswani, Shazeer, Parmar, Uszkoreit, Jones,
  Gomez, Kaiser, and Polosukhin}{Vaswani et~al\mbox{.}}{2017}]%
        {vaswani2017attention}
\bibfield{author}{\bibinfo{person}{Ashish Vaswani}, \bibinfo{person}{Noam
  Shazeer}, \bibinfo{person}{Niki Parmar}, \bibinfo{person}{Jakob Uszkoreit},
  \bibinfo{person}{Llion Jones}, \bibinfo{person}{Aidan~N Gomez},
  \bibinfo{person}{{\L}ukasz Kaiser}, {and} \bibinfo{person}{Illia
  Polosukhin}.} \bibinfo{year}{2017}\natexlab{}.
\newblock \showarticletitle{Attention is all you need}. In
  \bibinfo{booktitle}{\emph{Advances in neural information processing
  systems}}. \bibinfo{pages}{5998--6008}.
\newblock


\bibitem[\protect\citeauthoryear{Verma and Ganguly}{Verma and Ganguly}{2019a}]%
        {10.1145/3331184.3331377}
\bibfield{author}{\bibinfo{person}{Manisha Verma} {and}
  \bibinfo{person}{Debasis Ganguly}.} \bibinfo{year}{2019}\natexlab{a}.
\newblock \showarticletitle{LIRME: Locally Interpretable Ranking Model
  Explanation}. In \bibinfo{booktitle}{\emph{Proceedings of the 42nd
  International ACM SIGIR Conference on Research and Development in Information
  Retrieval}} (Paris, France) \emph{(\bibinfo{series}{SIGIR'19})}.
  \bibinfo{publisher}{Association for Computing Machinery},
  \bibinfo{address}{New York, NY, USA}, \bibinfo{pages}{1281–1284}.
\newblock
\showISBNx{9781450361729}
\urldef\tempurl%
\url{https://doi.org/10.1145/3331184.3331377}
\showDOI{\tempurl}


\bibitem[\protect\citeauthoryear{Verma and Ganguly}{Verma and Ganguly}{2019b}]%
        {verma2019lirme}
\bibfield{author}{\bibinfo{person}{Manisha Verma} {and}
  \bibinfo{person}{Debasis Ganguly}.} \bibinfo{year}{2019}\natexlab{b}.
\newblock \showarticletitle{LIRME: Locally Interpretable Ranking Model
  Explanation}. In \bibinfo{booktitle}{\emph{Proceedings of the 42nd
  International ACM SIGIR Conference on Research and Development in Information
  Retrieval}}. ACM, \bibinfo{pages}{1281--1284}.
\newblock


\bibitem[\protect\citeauthoryear{Viera, Garrett, et~al\mbox{.}}{Viera
  et~al\mbox{.}}{2005}]%
        {viera2005understanding}
\bibfield{author}{\bibinfo{person}{Anthony~J Viera}, \bibinfo{person}{Joanne~M
  Garrett}, {et~al\mbox{.}}} \bibinfo{year}{2005}\natexlab{}.
\newblock \showarticletitle{Understanding interobserver agreement: the kappa
  statistic}.
\newblock \bibinfo{journal}{\emph{Fam med}} \bibinfo{volume}{37},
  \bibinfo{number}{5} (\bibinfo{year}{2005}), \bibinfo{pages}{360--363}.
\newblock


\bibitem[\protect\citeauthoryear{Wang, Wang, Jia, and Yin}{Wang
  et~al\mbox{.}}{2018}]%
        {wang2018explainable}
\bibfield{author}{\bibinfo{person}{Nan Wang}, \bibinfo{person}{Hongning Wang},
  \bibinfo{person}{Yiling Jia}, {and} \bibinfo{person}{Yue Yin}.}
  \bibinfo{year}{2018}\natexlab{}.
\newblock \showarticletitle{Explainable Recommendation via Multi-Task Learning
  in Opinionated Text Data}.
\newblock \bibinfo{journal}{\emph{SIGIR}} (\bibinfo{year}{2018}).
\newblock


\bibitem[\protect\citeauthoryear{Wang, Liu, Liu, Liu, Liu, and Mei}{Wang
  et~al\mbox{.}}{2020}]%
        {wang2020metasearch}
\bibfield{author}{\bibinfo{person}{Qi Wang}, \bibinfo{person}{Xinchen Liu},
  \bibinfo{person}{Wu Liu}, \bibinfo{person}{An-An Liu},
  \bibinfo{person}{Wenyin Liu}, {and} \bibinfo{person}{Tao Mei}.}
  \bibinfo{year}{2020}\natexlab{}.
\newblock \showarticletitle{MetaSearch: Incremental Product Search via Deep
  Meta-Learning}.
\newblock \bibinfo{journal}{\emph{IEEE Transactions on Image Processing}}
  \bibinfo{volume}{29} (\bibinfo{year}{2020}), \bibinfo{pages}{7549--7564}.
\newblock


\bibitem[\protect\citeauthoryear{Wiegreffe and Pinter}{Wiegreffe and
  Pinter}{2019}]%
        {wiegreffe2019attention}
\bibfield{author}{\bibinfo{person}{Sarah Wiegreffe} {and}
  \bibinfo{person}{Yuval Pinter}.} \bibinfo{year}{2019}\natexlab{}.
\newblock \showarticletitle{Attention is not not Explanation}.
\newblock \bibinfo{journal}{\emph{Proceddings of the 2019 Conference on
  Empirical Methods in Natural Language Processing}} (\bibinfo{year}{2019}).
\newblock


\bibitem[\protect\citeauthoryear{Wu, Yan, and Si}{Wu et~al\mbox{.}}{2017}]%
        {wu2017ensemble}
\bibfield{author}{\bibinfo{person}{Chen Wu}, \bibinfo{person}{Ming Yan}, {and}
  \bibinfo{person}{Luo Si}.} \bibinfo{year}{2017}\natexlab{}.
\newblock \showarticletitle{Ensemble methods for personalized e-commerce search
  challenge at CIKM Cup 2016}.
\newblock \bibinfo{journal}{\emph{arXiv preprint arXiv:1708.04479}}
  (\bibinfo{year}{2017}).
\newblock


\bibitem[\protect\citeauthoryear{Yosinski, Clune, Bengio, and Lipson}{Yosinski
  et~al\mbox{.}}{2014}]%
        {yosinski2014transferable}
\bibfield{author}{\bibinfo{person}{Jason Yosinski}, \bibinfo{person}{Jeff
  Clune}, \bibinfo{person}{Yoshua Bengio}, {and} \bibinfo{person}{Hod Lipson}.}
  \bibinfo{year}{2014}\natexlab{}.
\newblock \showarticletitle{How transferable are features in deep neural
  networks?}. In \bibinfo{booktitle}{\emph{Advances in neural information
  processing systems}}. \bibinfo{pages}{3320--3328}.
\newblock


\bibitem[\protect\citeauthoryear{Zamani and Croft}{Zamani and Croft}{2020}]%
        {zamani2020learning}
\bibfield{author}{\bibinfo{person}{Hamed Zamani} {and} \bibinfo{person}{W~Bruce
  Croft}.} \bibinfo{year}{2020}\natexlab{}.
\newblock \showarticletitle{Learning a Joint Search and Recommendation Model
  from User-Item Interactions}. In \bibinfo{booktitle}{\emph{Proceedings of the
  13th International Conference on Web Search and Data Mining}}.
  \bibinfo{pages}{717--725}.
\newblock


\bibitem[\protect\citeauthoryear{Zeiler and Fergus}{Zeiler and Fergus}{2014}]%
        {zeiler2014visualizing}
\bibfield{author}{\bibinfo{person}{Matthew~D Zeiler} {and} \bibinfo{person}{Rob
  Fergus}.} \bibinfo{year}{2014}\natexlab{}.
\newblock \showarticletitle{Visualizing and understanding convolutional
  networks}. In \bibinfo{booktitle}{\emph{European conference on computer
  vision}}. Springer, \bibinfo{pages}{818--833}.
\newblock


\bibitem[\protect\citeauthoryear{Zhang}{Zhang}{2016}]%
        {zhang2016explainable}
\bibfield{author}{\bibinfo{person}{Yongfeng Zhang}.}
  \bibinfo{year}{2016}\natexlab{}.
\newblock \showarticletitle{Explainable Recommendation: Theory and
  Applications}.
\newblock \bibinfo{journal}{\emph{PhD thesis}} (\bibinfo{year}{2016}).
\newblock


\bibitem[\protect\citeauthoryear{Zhang and Chen}{Zhang and Chen}{2020}]%
        {zhang2020explainable}
\bibfield{author}{\bibinfo{person}{Yongfeng Zhang} {and} \bibinfo{person}{Xu
  Chen}.} \bibinfo{year}{2020}\natexlab{}.
\newblock \showarticletitle{Explainable recommendation: A survey and new
  perspectives}.
\newblock \bibinfo{journal}{\emph{Foundations and Trends in Information
  Retrieval}} \bibinfo{volume}{14}, \bibinfo{number}{1} (\bibinfo{year}{2020}),
  \bibinfo{pages}{1--101}.
\newblock


\bibitem[\protect\citeauthoryear{Zhang, Lai, Zhang, Zhang, Liu, and Ma}{Zhang
  et~al\mbox{.}}{2014}]%
        {zhang2014explicit}
\bibfield{author}{\bibinfo{person}{Yongfeng Zhang}, \bibinfo{person}{Guokun
  Lai}, \bibinfo{person}{Min Zhang}, \bibinfo{person}{Yi Zhang},
  \bibinfo{person}{Yiqun Liu}, {and} \bibinfo{person}{Shaoping Ma}.}
  \bibinfo{year}{2014}\natexlab{}.
\newblock \showarticletitle{Explicit factor models for explainable
  recommendation based on phrase-level sentiment analysis}. In
  \bibinfo{booktitle}{\emph{Proceedings of the 37th international ACM SIGIR
  conference on Research \& development in information retrieval}}. ACM,
  \bibinfo{pages}{83--92}.
\newblock


\bibitem[\protect\citeauthoryear{Zilke, Menc{\'\i}a, and Janssen}{Zilke
  et~al\mbox{.}}{2016}]%
        {zilke2016deepred}
\bibfield{author}{\bibinfo{person}{Jan~Ruben Zilke},
  \bibinfo{person}{Eneldo~Loza Menc{\'\i}a}, {and} \bibinfo{person}{Frederik
  Janssen}.} \bibinfo{year}{2016}\natexlab{}.
\newblock \showarticletitle{DeepRED--Rule extraction from deep neural
  networks}. In \bibinfo{booktitle}{\emph{International Conference on Discovery
  Science}}. Springer, \bibinfo{pages}{457--473}.
\newblock


\end{thebibliography}

\end{document}